\documentclass[
superscriptaddress,
preprint,
bibnotes,
longbibliography,
amsmath,
amssymb
]{revtex4-2}

\usepackage{graphicx}
\usepackage{bm}
\usepackage[hidelinks]{hyperref}
\usepackage{cleveref}
\usepackage{parskip}

\begin{document}

\title{Multi-scale time-resolved electron diffraction: A case study in moir\'e materials}

\author{C.J.R. Duncan}
\affiliation{Cornell Laboratory for Accelerator-Based Sciences and Education, Cornell University, Ithaca, NY 14850,  USA}
\author{M. Kaemingk}
\affiliation{Cornell Laboratory for Accelerator-Based Sciences and Education, Cornell University, Ithaca, NY 14850,  USA}
\author{W.H. Li}
\affiliation{Cornell Laboratory for Accelerator-Based Sciences and Education, Cornell University, Ithaca, NY 14850,  USA}
\author{M.B. Andorf}
\affiliation{Cornell Laboratory for Accelerator-Based Sciences and Education, Cornell University, Ithaca, NY 14850,  USA}
\author{A.C. Bartnik}
\affiliation{Cornell Laboratory for Accelerator-Based Sciences and Education, Cornell University, Ithaca, NY 14850,  USA}
\author{A. Galdi}
\affiliation{Cornell Laboratory for Accelerator-Based Sciences and Education, Cornell University, Ithaca, NY 14850,  USA}
\author{M. Gordon}
\affiliation{Cornell Laboratory for Accelerator-Based Sciences and Education, Cornell University, Ithaca, NY 14850,  USA}
\author{C.A. Pennington}
\affiliation{Cornell Laboratory for Accelerator-Based Sciences and Education, Cornell University, Ithaca, NY 14850,  USA}
\author{I.V. Bazarov}
\affiliation{Cornell Laboratory for Accelerator-Based Sciences and Education, Cornell University, Ithaca, NY 14850,  USA}
\author{H.J. Zeng}
\affiliation{Department of Chemistry, Stanford University, Stanford, CA 94305,  USA}
\author{F. Liu}
\affiliation{Department of Chemistry, Stanford University, Stanford, CA 94305,  USA}
\author{D. Luo}
\affiliation{SLAC National Accelerator Laboratory, Menlo Park, CA 94205,  USA}
\author{A. Sood}
\affiliation{Department of Mechanical and Aerospace Engineering, Princeton University, Princeton, NJ 08540,  USA}
\affiliation{Princeton Materials Institute, Princeton University, Princeton, NJ, 08540,  USA}
\author{A.M. Lindenberg}
\affiliation{Department of Materials Science and Engineering, Stanford University, Stanford, CA, 94305, USA}
\author{M.W. Tate}
\affiliation{Laboratory of Atomic and Solid State Physics, Cornell University, Ithaca, NY, 14853, USA}
\author{D.A. Muller}
\affiliation{Kavli Institute at Cornell for Nanoscale Science, Ithaca, NY 14853, USA}
\affiliation{School of Applied and Engineering Physics, Cornell University, Ithaca, NY 14853 USA}
\author{J. Thom-Levy}
\affiliation{Laboratory for Elementary-Particle Physics, Cornell University, Ithaca, NY 14853, USA}
\author{S.M. Gruner}
\affiliation{Laboratory of Atomic and Solid State Physics, Cornell University, Ithaca, NY, 14853, USA}
\affiliation{Kavli Institute at Cornell for Nanoscale Science, Ithaca, NY 14853, USA}
\author{J.M. Maxson}
\affiliation{Cornell Laboratory for Accelerator-Based Sciences and Education, Cornell University, Ithaca, NY 14850,  USA}

\begin{abstract}
Ultrafast-optical-pump --- structural-probe measurements, including ultrafast electron and x-ray scattering, provide direct experimental access to the fundamental timescales of atomic motion, and are thus
foundational techniques for studying matter out of equilibrium. High-performance detectors are needed in scattering experiments to obtain maximum scientific value from every probe particle. We deploy a hybrid pixel array direct electron detector to perform ultrafast electron diffraction experiments on a WSe$_2$/MoSe$_2$ 2D heterobilayer, resolving the weak features of diffuse scattering and moir\'e superlattice structure without saturating the zero order peak. Enabled by the detector's high frame rate, we show that a chopping technique provides diffraction difference images with signal-to-noise at the shot noise limit. Finally, we demonstrate that a fast detector frame rate coupled with a high repetition rate probe can provide continuous time resolution from femtoseconds to seconds, enabling us to perform a scanning ultrafast electron diffraction experiment that maps thermal transport in WSe$_2$/MoSe$_2$ and resolves distinct diffusion mechanisms in space and time.
\end{abstract}

\maketitle

\section{Introduction}

Ultrafast x-ray and electron scattering experiments are essential tools in the
materials-by-design thrust of modern condensed matter physics and engineering,
as they can probe far-from-equilibrium dynamic phenomena at atomic space and
time
scales~\cite{ihee2001direct,cavalleri2001femtosecond,siwick2003atomic,gedik2007nonequilibrium,stojchevska2014ultrafast,sie_ultrafast_2019}.
In this discipline, technological improvements in probe quality (space/momentum
and time/energy resolution) and probe detection (sensitivity, speed) are key to
enabling the discovery of new non-equilibrium material
functionality~\cite{broholm2016basic,carini2012neutron}.   

Atomically thin moir\'e materials, such as twisted homobilayers or
heterobilayers~\cite{cao2018unconventional,yoo2019atomic,liao2020precise,liu2020disassembling,kim2021extremely,zhao2021high,gadelha2021localization},
present exciting opportunities for ultrafast study given the wide array of
\emph{equilibrium} quantum phenomena observed in these systems, including
superconductivity and orbital magnetism, that are \emph{tunable} with twist
angle~\cite{basov2017towards,andrei2021marvels}. Appropriately tuned ultrafast
excitation may be a route to optical switching of these properties, as has been
observed in other quantum materials~\cite{fausti2011light,sie_ultrafast_2019}.

Many of the properties that make 2D materials physically interesting
also pose significant challenges for ultrafast structural probes. Atomically
thin films are inherently weak scatterers and this challenge is compounded by
the fact that compact ultrafast probes generally offer lower time-average flux
than their non-time-resolved
counterparts~\cite{weathersby2015mega,feist2017ultrafast,Li2020HHG}. Further,
important features in the information-rich diffraction patterns from these
materials can be separated by many orders of magnitude in intensity: descending
from the (0,0,0) peak, to Bragg scattering, to weaker satellite peaks caused by
longer wavelength periodic lattice distortions (PLD), to yet weaker thermal
diffuse scattering (TDS)~\cite{britt2022direct}. In pump--probe ultrafast
electron diffraction (UED), we seek to measure small changes in this already
weak scattering. For example, important details of electron--phonon coupling
can be found in parts per thousand modulation of scattering signals that in
static diffraction are already at the $10^{-4}$ level of total beam current.
Moir\'e materials exemplify this challenge, and could plausibly present
correlated signatures of interlayer interactions in Bragg, PLD and TDS
scattering, with tunable superstructure periodicity that can extend to tens of
$\mathrm{nm}$~\cite{yoo2019atomic}. 

\begin{figure}
\includegraphics[width=0.8\columnwidth]{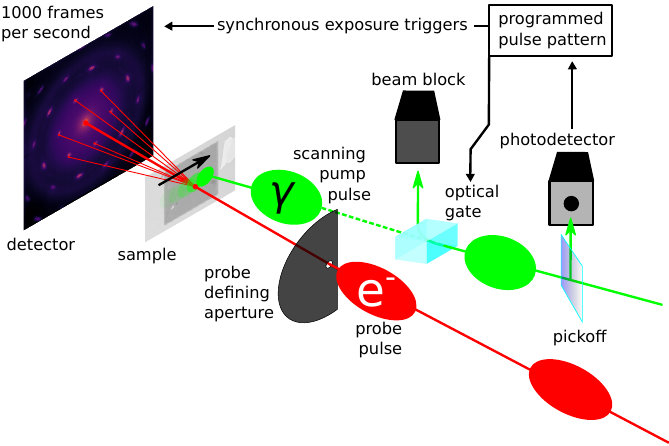}
\caption{Schematic of our ultrafast electron diffraction setup, which enables our femtosecond, micron-sized probe to measure both the spatial and temporal dependence of the sample response to pumping with femtosecond, $515 \ \mathrm{nm}$ laser pulses. Pump pulses are independently gated, synchronously with detector triggering, allowing us to isolate individual probe pulses arriving from $10^{-13}$ to $10^{-3}$ s after the pump pulse.\label{fig:timing}}
\end{figure}

To meet the material-science need to investigate ultrafast evolution of
diffraction features  across multiple intensity and length scales
simultaneously -- in the same scattering data set and with Poisson-limited
experimental precision -- requires bright probe sources and detectors with
single-particle sensitivity and high dynamic range. Workhorse indirect
detectors function by detecting light generated in a scintillator, which places
limits on dynamic range, spatial resolution, and
sensitivity~\cite{meyer1998effects,fan2000digital,zuo2000electron,gruner2002charge}.
The standard metric for detector performance is detective quantum efficiency
(DQE): state-of-the-art indirect detectors offer DQE at best in the few tens of
percent, whereas a purely shot noise limited detection system would have a DQE
of unity~\cite{ruskin2013quantitative}.
Among devices that aim to overcome these
limitations~\cite{vecchione2017direct}, direct detectors are now an established
tool at x-ray user
facilities~\cite{allahgholi2019megapixels,leonarski2018fast,tate2013medium},
and enable computational imaging with state-of-the-art resolution in
non-time-resolved electron
microscopy~\cite{jiang2018electron,chen2021electron}, higher energy resolution
in electron-energy-loss spectroscopy~\cite{hart2017}, and higher
spatial-resolution in electron cryo-microscopy~\cite{shen2018}. 

The most common approach to direct electron detection is \textit{pulse
counting}~\cite{faruqi2007electronic}, which saturates at one particle per
pixel per relaxation period (order 100~ns) and thus cannot handle the large
peak currents typical of UED experiments~\cite{weathersby2015mega}. Integrating
direct detectors, by comparison, accurately report the total probe-particle
energy incident on each pixel and are uniquely suited for applications with
high-intensity pulsed beams of sub-$\mathrm{ps}$ duration, such as
UED~\cite{lee2017ultrafast}. 

The direct electron detector we employ in this work, the Electron Microscope
Pixel Array Detector (EMPAD) has a DQE (at zero spatial frequency) of 0.95,
with a signal to noise ratio for individual electron detection approaching 100
for 140 keV electrons. Its novel in-pixel, hybrid analog/digital circuitry
simultaneously provides very high dynamic range, as described
below~\cite{tate2016high}. 

Fluctuations in the probe current incident on the sample also limit precision when measuring scattering rates, even if the detector counts scattered particles perfectly. Causes include variation in laser power on the photocathode, changes in the quantum efficiency of the photocathode, and -- critical in micro-diffraction --  unavoidable jitter and drift in the probe beam position on the probe defining aperture.  Fluctuations in probe current fail to oscillate around a time-independent mean: thus, experimental precision does not improve from simply averaging longer acquisitions.

Instead, the incident probe current often exhibits universal $1/f$
frequency dependence. The impact is severe when probing both real-space and
reciprocal space features of the sample, as done in micro-diffraction
experiments~\cite{li2022kiloelectron,shen2018femtosecond}.
In optical pump--probe modalities, such as absorption spectroscopy,
high-frequency pulse chopping and lock-in detection is an essential technique
that reduces frequency-dependent noise~\cite{schriever2008ultrasensitive}, but
impossible to adopt in UED without a fast frame rate detector capable of
measuring the reference signal at frequencies well above those of the noise
sources. 

Ultimately, sample lifetime under the stress of repeated pump--probe cycles imposes a physical bound on total experiment time. Therefore, reducing acquisition time by eliminating sources of noise (other than fundamental Poisson uncertainty) significantly increases the breadth of feasible pump--probe experiments.

Here we deploy the EMPAD integrating electron detector, with our high
brightness keV ultrafast electron micro-diffraction
beamline~\cite{li2022kiloelectron}, to probe the out-of-equilibrium dynamics of
a WSe$_2$/MoSe$_2$ moir\'e bilayer at long spatial periodicity
(high resolution in reciprocal space) and ultrafast time
scales~\cite{tate2016high,li2022kiloelectron,bai2020excitons,wang2021diffusivity}.
For the first time in UED, we resolve  a $10 \ \mathrm{nm}$ periodic moir\'e
superlattice~\cite{adrian2016complete}. We integrate the superlattice signal
without saturating the more intense Bragg peaks caused by angstrom scale
interatomic spacing.  We demonstrate that the $1 \ \mathrm{kHz}$ detector frame rate
enables a fast pulse chopping technique that drastically improves
signal-to-noise in measurements of the sample response to ultrafast pumping. 

The fast $1 \ \mathrm{kHz}$ detector frame rate also enables a novel pulse-picking
technique, which extends the range of timescales accessible in ultrafast
pump--probe experiments beyond $\mathrm{\mu s}$ with $\mathrm{fs}$ precision.
Implementations of pump--probe delays at the $\mathrm{\mu s}$ scale typically rely on
electrical triggering, e.g.,~nanosecond Q-switched lasers or gain-switched
diodes with picosecond pulse duration~\cite{domke2012ultrafast}. Femtosecond
pulses, by increasing the peak laser field strength for the same deposited
energy, have the potential to unlock interesting metastable
behavior~\cite{disa2020polarizing}, and our method could resolve, e.g.,~$\mathrm{THz}$ frequency modulations $\mathrm{\mu s}$ after excitation. We demonstrate
our pulse-picking technique with a micro-diffraction probe and map the diffusion
of heat in our sample in space and time from initial $\mathrm{fs}$ ultrafast
excitation, out to $\mathrm{ms}$ thermal relaxation, with $\mathrm{\mu m}$ spatial
resolution.  Experimental access to $100 \ \mathrm{\mu s}$ timescales allows us to extract
from our space-and-time-resolved data the thermal diffusivity of our moir\'e
sample.

\section{Results}
\subsection{Resolving moir\'e superstructure}

\begin{figure}
\includegraphics[width=0.7\columnwidth]{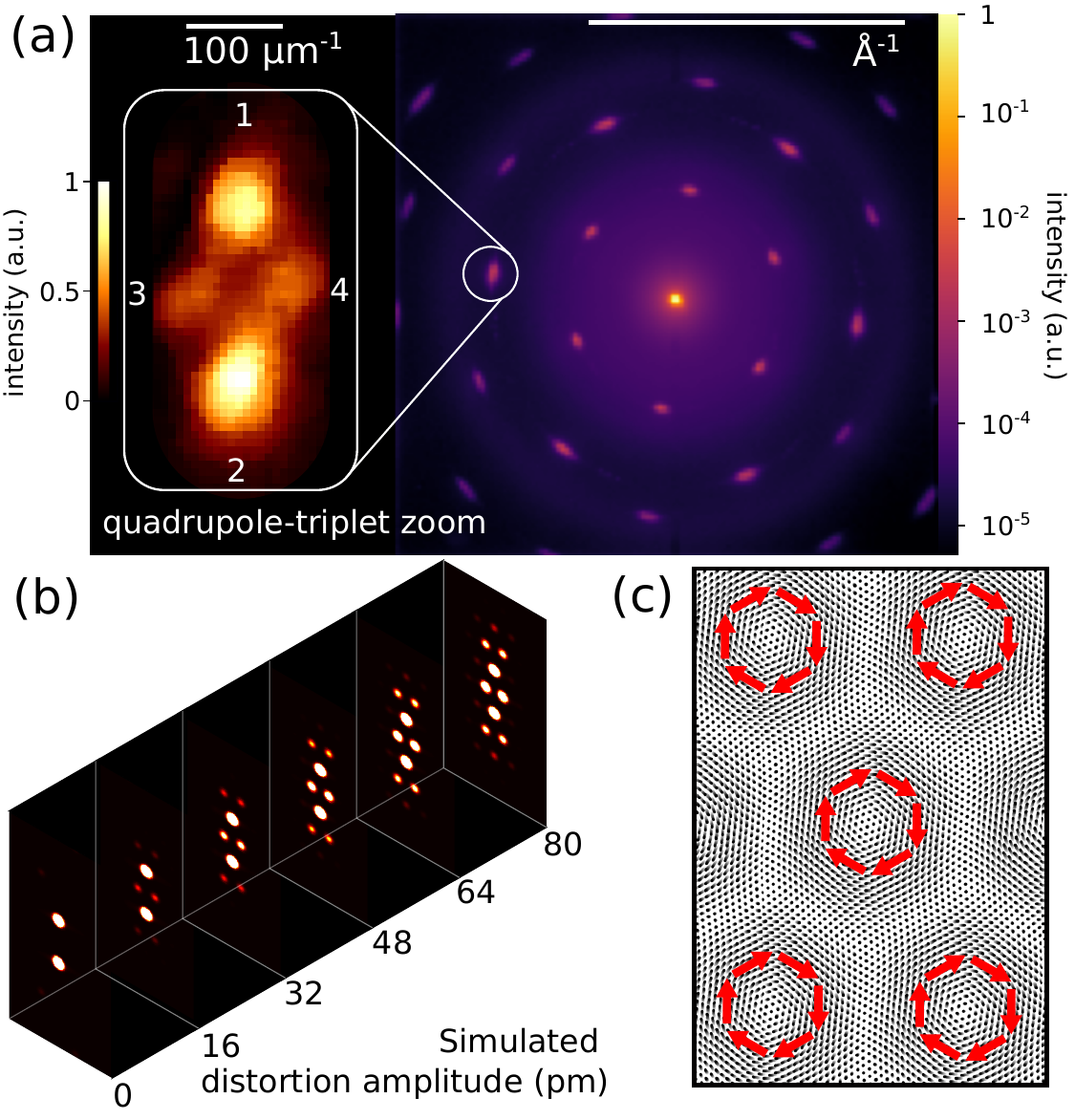}
\caption{(a) Logarithmic scale diffraction pattern obtained from $\mathrm{WSe}_2/\mathrm{MoSe}_2$ at short-camera length, demonstrating the EMPAD dynamic range. Inset left, a long camera-length diffraction difference pattern showing a single pair of moir\'e Bragg peaks, labeled 1 and 2: also visible, a pair of satellite peaks, labeled 3 and 4, with $10 \ \mathrm{nm}$ periodicity. This  inset image shows the absolute value of the difference in counts between a laser-pumped and unpumped sample at a delay of 4 ps: pumping enhances the contrast between the satellite peaks and the tails of the main Bragg peaks. The effect of beam chopping on the experimental uncertainty of peaks labeled 1 and 2 is shown in  Fig.~\ref{fig:shot}. Detector camera length is varied with a magnetic quadrupole lens triplet. At the camera length shown inset, the width of one detector pixel is $2.6\times10^{-3}$ reciprocal lattice units (rlu) and the electron beam spot size on the detector is $1.1\times 10^{-2}$ rlu, where $1$ rlu $:=  0.30~\text{\AA}^{-1}$. (b) Simulated diffraction patterns as a function of period-lattice-distortion (PLD) amplitude; PLD is driven by interlayer interaction between $\mathrm{WSe}_2$ and $\mathrm{MoSe}_2$. (c) Simulated moir\'e lattice visualized in real space; red arrows indicate direction of local atomic displacement.\label{fig:diffraction}}
\end{figure}

Our beamline setup is summarized in  Fig.~\ref{fig:timing} and Appendix~\ref{methods};
additional details can be found in a previous work~\cite{li2022kiloelectron}.
To take best advantage of the detector frame rate, we independently gate pump
and probe pulse trains obtained from a femtosecond fiber-laser source with a
repetition rate of $125 \ \mathrm{kHz}$. The pump wavelength in the experiments reported
here is $515 \ \mathrm{nm}$. In micro-diffraction mode, depending on the beam coherence
required to resolve the reciprocal space structure of interest, typical charges
per pulse on target range from 10 to 1000 electrons.

 A magnetic quadrupole lens triplet controls the angular magnification of the diffraction pattern on the detector. Lens settings are summarized by \textit{camera length}: the hypothetical drift distance to the detector plane that would result in equivalent diffraction data.
 
  Figure~\ref{fig:diffraction}(a) shows a static diffraction pattern obtained from the $\mathrm{WSe}_2/ \mathrm{MoSe}_2$ sample at short camera length. The six-fold symmetry of the diffraction pattern arises from the underlying symmetry of the $\mathrm{WSe}_2/ \mathrm{MoSe}_2$ lattices. The logarithmic scale color bar includes over four orders of magnitude of contrast from the (0,0,0) peak to thermal diffuse scattering. 
 
 The EMPAD consists of a grid of $128 \ \times \ 128$ pixels each 150 $\mathrm{\mu m}$ $\times \ 150$
$\mathrm{\mu m}$ in size. Each pixel has a $500~\mathrm{\mu m}$ thick reverse-biased silicon
diode bump bonded to its own read-out electronic circuit. The well of each
pixel can record up to $10^6$ electron incidents per exposure, but the
detector saturates at lower counts if the rate of incidents exceeds a
threshold. Previous measurements performed with continuous beam illumination
(\textit{cw}) estimated this threshold rate to be $22$ electrons per
pixel per microsecond at our $140 \ \mathrm{keV}$ beam energy. Surprisingly, in our
pulsed experiments we observe saturation at the significantly higher rate of 60
electrons per pixel per sub-picosecond pulse, a result likely analogous to
effects studied in the context of x-ray free electron laser applications, and
discussed further in Appendix~\ref{saturation}. The EMPAD design is under active
development and the latest iteration (not deployed in our experiments)
increases the cw saturation level to $10^3$ electrons per pixel per
microsecond~\cite{philipp2022very}.
 
 The inset to  Fig.~\ref{fig:diffraction}(a) shows a long-camera-length diffraction difference pattern isolating a single pair of Bragg peaks $(\bar{1}\bar{1}20)$, aligned vertically in the image, with two additional satellite peaks visible, aligned horizontally. The difference pattern is formed by subtracting pumped exposures from unpumped at a delay of 4 ps. In an unpumped diffraction pattern, the more intense of the two highlighted Bragg peaks results from scattering off the $\mathrm{WSe}_2$  monolayer, the other from the $\mathrm{MoSe}_2$ monolayer. The reverse is true of the pumped difference image inset to  Fig.~\ref{fig:diffraction}(a): the more intense peak results from the stronger response of the $\mathrm{MoSe}_2$ layer. The interlayer twist angle controls the separation in reciprocal space between the two Bragg peaks,  measured here to be $2^\circ$. 
 
 The sixfold symmetry of the real-space moir\'e pattern entails that scattering
from the moir\'e superlattice forms a hexagon dressing each Bragg peak. The
satellites observable in the inset to  Fig.~\ref{fig:diffraction}(a) lie at
scattering vectors where monolayer contributions overlap. Static selected area
electron diffraction (SAED) from twisted bilayer graphene performed by others
shows that the intensity of moir\'e satellite peaks decays exponentially with
twist angle between $0.5^\circ$--$2^\circ$~\cite{yoo2019atomic}. These
results, and static SAED data from $1^\circ$ twisted
$\mathrm{WSe}_2/\mathrm{MoS}_2$~\cite{kim2022anomalous}, are consistent with our observation of two
satellites along the midline of a $(\bar{1}\bar{1}20)$ Bragg pair. Interlayer
interactions strain the monolayers: simulated diffraction patterns presented in
Fig.~\ref{fig:diffraction}(b) show that satellite peaks are absent without this
strain and grow in intensity as strain increases (see Appendix~\ref{simulation} for
simulation details). A future work will present time-series data showing the
detailed dynamics of the moir\'e superstructure.
 
\subsection{Reaching Poisson-limited experimental uncertainty}
 
\begin{figure}
\includegraphics[width=0.8\columnwidth]{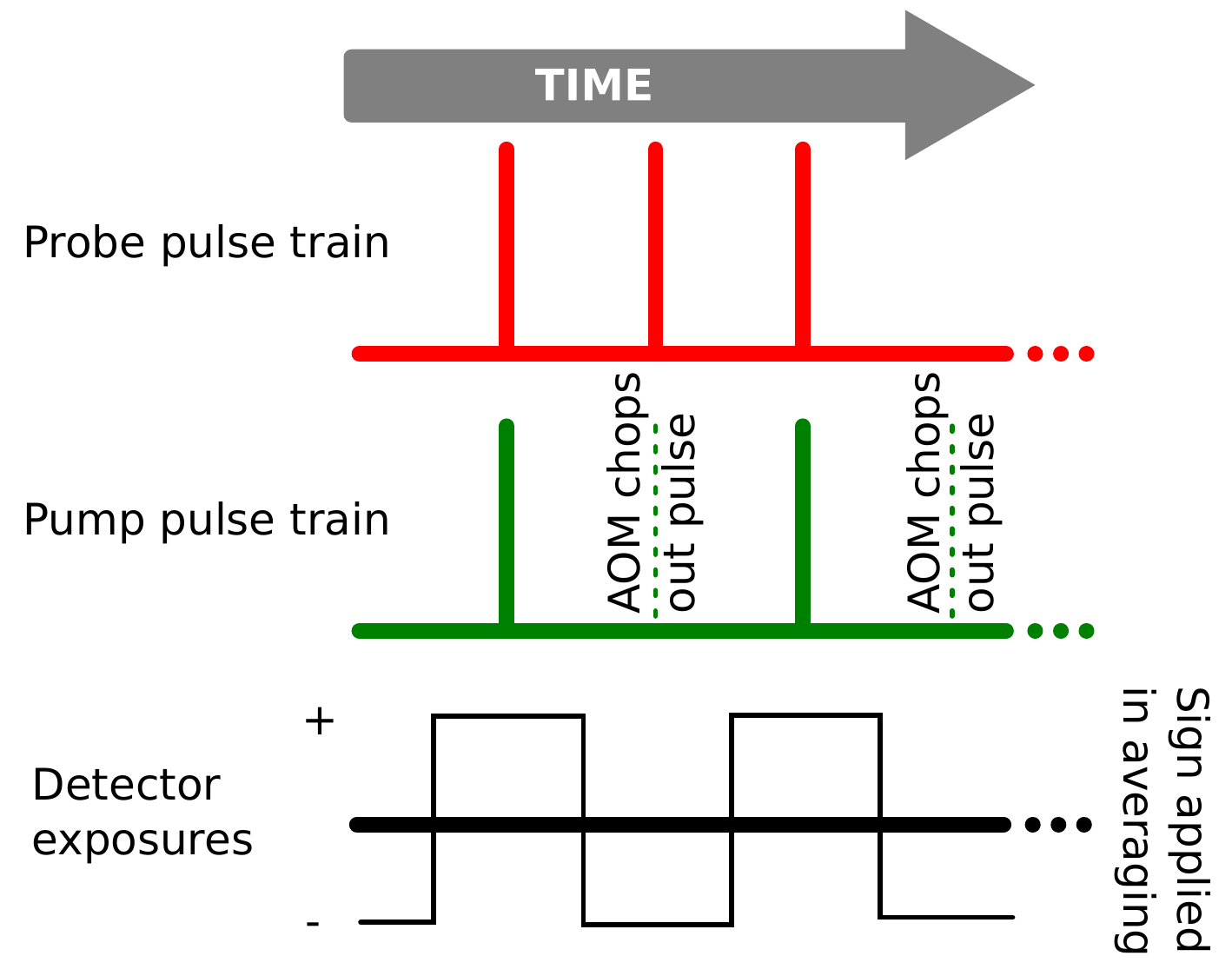}
\caption{Overview of beam chopping scheme: the probe pulses shown in red trigger an acousto-optic modulator (AOM, see  Fig.~\ref{fig:timing}) to block pump pulses, shown in green, from reaching the sample. Probe pulses synchronously trigger detector exposures, shown in black. Exposures enter the ensemble average with a positive sign if the pump gate is open, negative if the pump gate is blocked.\label{fig:chopping}}
\end{figure}

In this section, we demonstrate with experimental data how fast beam chopping coupled with high detector frame rate dramatically improves signal to noise in ultrafast pump--probe experiments. The chopping concept is illustrated in  Fig.~\ref{fig:chopping}. To put the results of this section in context, it is necessary to briefly summarize the relevant sources of noise. The typical signal in UED is the fractional change in scattering intensity $\Delta I/ I$ as a function of delay time and scattering angle. Distinguishing \textit{pumped} exposures, when the pump beam is incident on the sample, from \textit{unpumped} exposures, when the pump beam is off, $\Delta I/ I$ can be expressed in terms of the ratio of the number $N_{\mathrm{s}}$ of scattered electrons within a given solid angle to the number $N_{\mathrm{i}}$ of incident electrons:
\begin{equation}
\label{observable}
    \frac{\Delta I}{I} = \frac{\left(N_{sp}/N_{ip} \right)}{\left( N_{su}/N_{iu} \right)} - 1,
\end{equation}
 where the additional subscript $p$ or $u$
corresponds to the pumped or unpumped condition. The contribution of the probe
beam to total experimental uncertainty $\sigma_{\mathrm{total}}$ in estimating $\Delta I/I$ is
then analyzed by considering contributions from each $N$ appearing in
Eq.~\eqref{observable}. These contributions are conveniently aggregated into
four terms~\cite{kealhofer2015signal}:
\begin{equation}
\label{uncertainty_budget}
\sigma^2_{\mathrm{total}} = \sigma^2_{\mathrm{shot}} +  \sigma^2_{\mathrm{detector}} + \sigma^2_{\mathrm{transport}} + \sigma^2_{\mathrm{source}}.
\end{equation}
Each term in Eq.~\eqref{uncertainty_budget} corresponds to elements that
together comprise the entire scattering experiment: $\sigma_{\mathrm{shot}}$ accounts for
the Poisson distribution in the number of electron scattering events;
$\sigma_{\mathrm{detector}}$ accounts for uncertainty introduced in converting electron
incidence into detector counts; finally, $\sigma_{\mathrm{source}}$ and $\sigma_{\mathrm{transport}}$ account for
uncertainty introduced in emitting and transporting electrons from source to
detector. As the relative importance of each term depends on signal
intensity~\cite{kealhofer2015signal}, all four can be important in an
experiment that aims to resolve diffraction features at separated intensity
scales  simultaneously.

The  shot noise contribution $\sigma_{\mathrm{shot}}$ to $\Delta I/ I$ is dominated by the contribution from scattered electrons, so that:
\begin{equation}
    \sigma^2_{\mathrm{shot}} = \left(1 + \frac{\Delta I}{I}\right)^2\left(\frac{1}{N_{sp}} + \frac{1}{N_{su}} \right).
\end{equation}
As a numerical illustration, to resolve a 1\% change in a diffraction feature
intensity with 0.5\% rms uncertainty requires a minimum of $\sim10^5$
electrons scattered into that feature. The corresponding minimum accumulation
time, supposing a high charge machine with $10^6$ electrons per pulse at a
$1 \ \mathrm{kHz}$ repetition rate~\cite{britt2022direct}, and $10^{-3}$ scattering
factor, is only $100 \ \mathrm{ms}$. Nonetheless, stroboscopic UED experiments often
integrate for much longer to achieve this same level of experimental
uncertainty
(e.g.,~\cite{britt2022direct,chase2016ultrafast,yang2020simultaneous}),
because in these experiments $\sigma^2_{\mathrm{shot}}$ is not the most significant term in
$\sigma^2_{\mathrm{total}}$.

Contributions to $\sigma_{\mathrm{detector}}$ depend on the detector technology, and include per
count amplification noise as well as readout noise~\cite{kealhofer2015signal}.
With respect to the EMPAD, rms read noise (where the average is taken over all
pixels) is equivalent to 0.011 electrons at $140 \ \mathrm{keV}$, a signal-to-noise of
100.~\cite{weiss2017high}. 

The final two terms in Eq.~\eqref{uncertainty_budget}, $\sigma_{\mathrm{transport}}$ and
$\sigma_{\mathrm{source}}$, are dominated by the contribution from incident electrons, which
arises from loss of information about the incident beam. Significant loss of
information occurs if the intensity of the beam at zero scattering angle is not
recorded. Failure to record this information can occur because the beam at zero
scattering angle saturates the dynamic range of the detector, but the EMPAD is
not saturated in the micro-diffraction experiments we report here.  Instead, in
our high-momentum-resolution experiments at long camera length, e.g.,~the inset
to  Fig.~\ref{fig:diffraction}(a), the 2~cm diameter of the $128 \ \times \ 128$ pixel
detector does not allow simultaneous sampling of the zero-angle peak together
with the diffraction features of interest  --- the sidebands due to moir\'e
interlayer interactions. This limitation could be overcome by combining
multiple detectors into a larger array, as has been demonstrated with a version
of the EMPAD in x-ray imaging~\cite{tate2013medium}. Jitter and drift arising
from electron transport are specific to the experimental setup and lab
environment. With reference to our setup, a micron scale probe defining
aperture in micro-diffraction mode makes the transmitted current sensitive (at
the 0.1\% -- \%1 level relevant in UED) to micro-radian changes in beam
pointing. Kicks of this size are likely the result of integrating
electromagnetic pollution (partially compensated by Helmholtz coils) along the
3~m distance from cathode to sample. Techniques to achieve electromagnetically,
as well as mechanically, cleaner lab environments developed for atomic
resolution scanning transmission microscopy could also be implemented to
improve beam transport in micro-UED~\cite{muller2006room}.

\begin{figure}
\includegraphics[width=0.8\columnwidth]{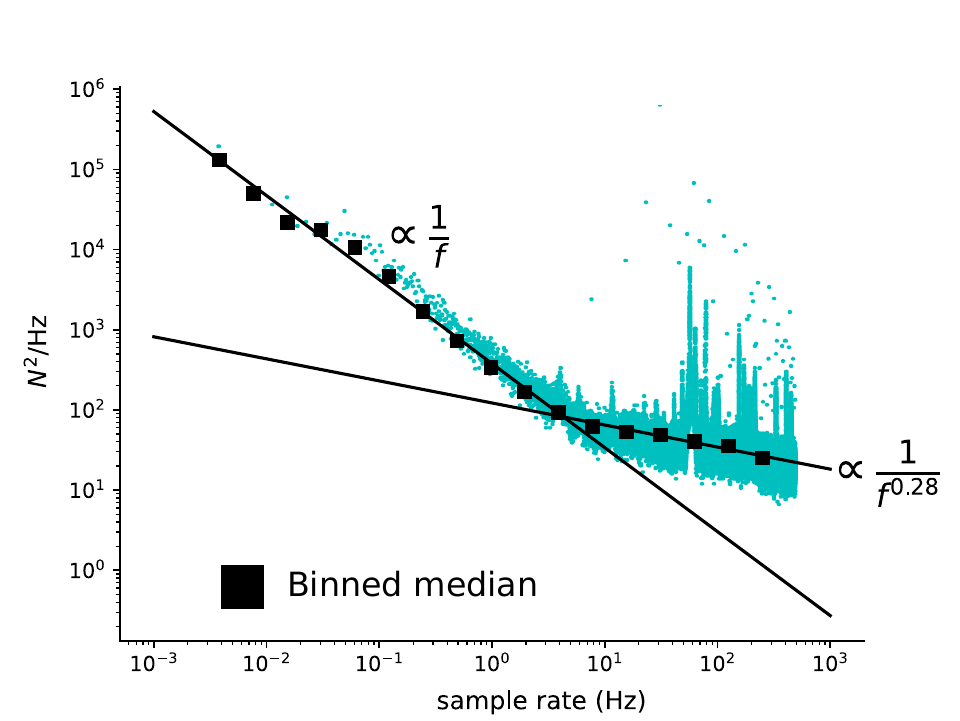}
\caption{Power spectral density (PSD) of the number of electrons transmitted through our 10 $\mathrm{\mu m}$ probe-defining aperture. Fluctuations in transmitted intensity are dominated by spatial drift of the beam on the probe-defining aperture. The PSD is computed from 90~min of time-series data using the Welch method. To fit the trend lines, the PSD is binned in exponentially spaced frequency intervals.\label{fig:psd}}
\end{figure}

Absent direct measurements,  the number of incident electrons must be estimated
from correlated information, for example, monitored photo-emission laser power
and acquisition time~\cite{kealhofer2015signal}. In our micro-diffraction
experiment, these indirect sources of information cannot track fluctuations due
to transport through our probe-defining aperture just upstream of the sample.
Figure~\ref{fig:psd} shows a measured frequency spectrum of total current transmitted
through our $10 \ \mathrm{\mu m}$ probe-defining aperture.

\begin{figure}
\includegraphics[width=0.8\columnwidth]{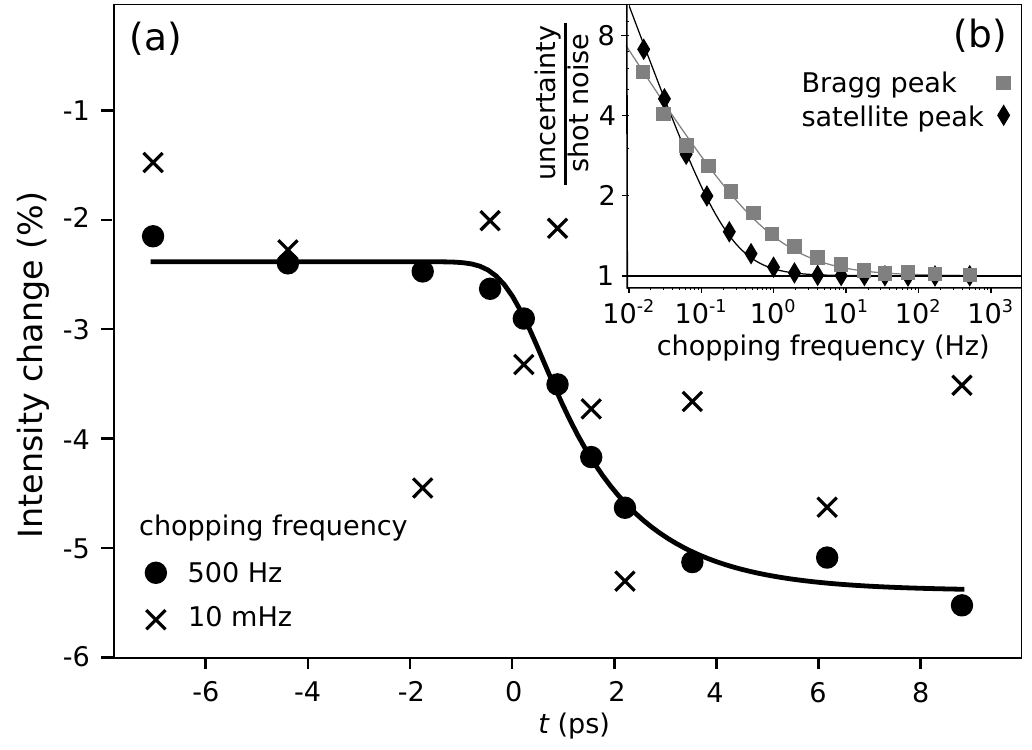}
\caption{(a) Measurement of the ultrafast Debye--Waller effect, summing peaks 1--4 in  Fig.~\ref{fig:diffraction}(a), comparing data quality with high and low chopping frequencies, each data point integrated for two minutes. (b) Measurement uncertainty as a function of chopping frequency: referring to the variance-to-mean ratio (VMR) defined in Eq.~\eqref{VMR}, the vertical axis shows $\sqrt{\mathrm{VMR}}$ and can be interpreted as the ratio of total uncertainty to the shot noise limit set by the total integration time. Two curves compare different scattering rates: gray squares show a scattering rate equivalent to the Bragg peak labeled 1 in  Fig.~\ref{fig:diffraction}(a), black diamonds show a scattering rate equivalent to the satellite peak labeled 4 in  Fig.~\ref{fig:diffraction}(a).\label{fig:shot}}
\end{figure}

To demonstrate that beam chopping eliminates jitter and drift as sources of
uncertainty in pump--probe experiments,  Fig.~\ref{fig:shot}(a) shows the
measurement of the ultrafast response at two chopping frequencies with the
small probe beam. The trend is an exponential decay convolved with the
instrument response~\cite{chatelain2014coherent}. The non-zero sample response
at delay times earlier than the pump arrival is due to $1 \ \mathrm{kHz}$ thermal
cycling, investigated below. Each data point is acquired with a two-minute
integration time. The spread in the data at the slow chopping rate (one minute
pumped, followed by one minute unpumped) is  comparable to the size of the
ultrafast effect. In stark contrast, the data acquired at the $500 \ \mathrm{Hz}$
chopping frequency follows the fitted trend closely. 

Further investigating the dependence of experimental uncertainty on beam-chopping frequency, we benchmark our experimental uncertainty with pump--probe time-series data, summarized in  Fig.~\ref{fig:shot}(b).
We can estimate frequency dependence by taking averages with the respect to an ensemble of time-bins of variable duration $W$. We compute a variance-to-mean ratio (VMR) as a function of $W$:
\begin{equation}
\label{VMR}
    \mathrm{VMR}(W) :=  \frac{\mathrm{Var}_{W}\left(N_{sp}-N_{su} \right)}{\mathrm{E}_{W}\left[ N_{sp}+N_{su}\right]},
\end{equation}
where $\mathrm{Var}_W(X)$ denotes the variance of $X$ over duration-$W$ time-bins and $\mathrm{E}_W[X]$ the mean over the same bins. The VMR defined in Eq.~\eqref{VMR} can be interpreted as the square of the experimental uncertainty in estimating the effect of pumping the sample, normalized by the shot noise limit. If $N_{su}, N_{sp}$ are drawn from independent Poisson distributions then the VMR  must equal one. The VMR rises above one with the introduction of fluctuations in the mean of the Poisson distribution, due, e.g.,~to fluctuating laser power on the photocathode, and $1/f$ noise in the steering magnets and accelerating voltage.

 Figure~\ref{fig:shot}(b) plots $\sqrt{\mathrm{VMR}}$ (as defined in Eq.~\eqref{VMR}) against chopping frequency and compares higher intensity Bragg scattering peaks with the 30\% weaker superlattice peaks. The two data sets are fit to power law trend lines. Both data-sets are at the shot noise floor when chopped at $500 \ \mathrm{Hz}$. The data reveals that the higher the scattering rate, the higher the chopping frequency required to hit the shot noise floor (holding fixed the total integration time). At the higher scattering rate, the VMR is appreciably greater than unity at chopping frequencies below $100 \ \mathrm{Hz}$, while at the lower scattering rate, the VMR begins to rise above shot noise only below $10 \ \mathrm{Hz}$. The explanation is that more counts for fixed acquisition time reduces the Poisson contribution to the noise in any given frequency band, increasing the experimental sensitivity to jitter and drift in the same band. It follows that for fixed acquisition time the chopping frequency required to reach the shot noise floor depends both on the experimental setup -- the average probe current -- and on the intrinsic details of the scattering mechanism, e.g.,~low intensity thermal diffuse scattering is shot noise limited at a lower chopping frequency than high intensity Bragg scattering.

\begin{figure}
\includegraphics[width=0.8\columnwidth]{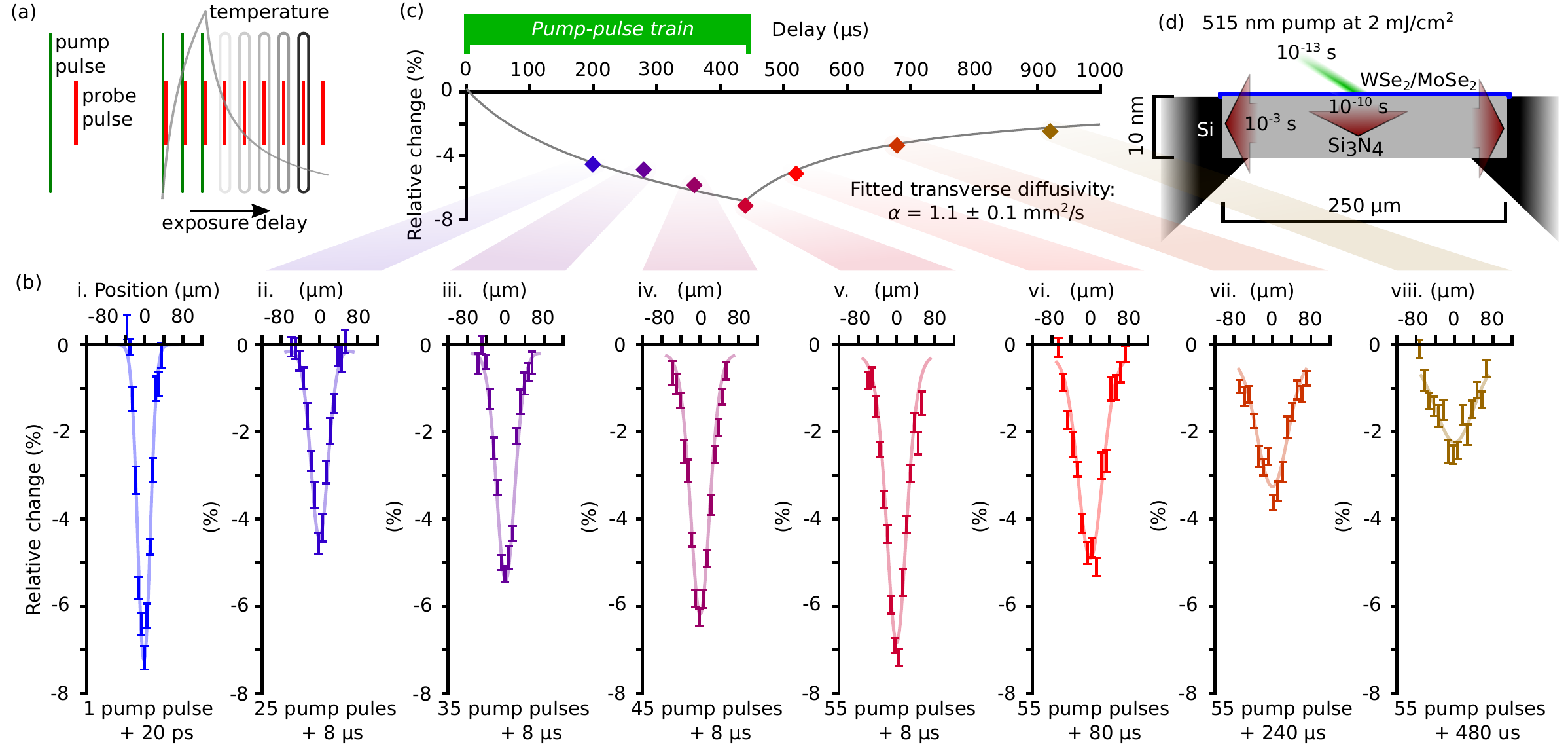}
\caption{Multi-dimensional, multi-scale sample thermometry. (a) Long-delay scanning technique: pump pulses are gated while microsecond exposures isolate individual probe pulses as they arrive at 8 $\mathrm{\mu s}$ intervals, sampling femtosecond pulses at delays up to a millisecond and longer. (b) i.--viii. Change in diffraction intensity $\Delta I/I$ (vertical axis) versus the spatial displacement of pump and probe (horizontal axis), at pump--probe delays spanning eight orders of magnitude. The diffraction signal is the sum of all electron counts in the highlighted region around the  $(\bar{1}\bar{1}20)$ peak indicated in  Fig.~\ref{fig:diffraction}: both monolayers contribute. (b) i. The ultrafast response $20 \ \mathrm{ps}$ after one pump pulse, with the solid line showing a Gaussian fit. (b) ii. The response $8 \ \mathrm{\mu s}$ after a $200 \ \mathrm{\mu s}$ pump-pulse train containing 25 pump pulses. (b) iii.-v. Extending the duration of the pump-pulse train to a maximum of $440 \ \mathrm{\mu s}$, the response $8 \ \mathrm{\mu s}$ after the final pulse in the train: the method is illustrated in 
 Fig.~\ref{fig:timing}. (b) vi.-viii. Sample relaxation, holding the $440 \ \mathrm{\mu s}$ pump-pulse train duration fixed. (c) $\Delta I/I$ (vertical axis), at $0 \ \mathrm{\mu m}$
spatial displacement, versus pump--probe delay. All solid lines in (b) ii.--viii. and (c) are cross sections of the fitted, three-parameter diffusion model summarized in Eq.~\eqref{heat}. The fitted transverse diffusivity
$\alpha = 1.1 \pm 0.1 \ \mathrm{mm}^2/{\rm s}$.
(d) Heat diffusion time-scales: the duration of pump excitation is $10^{-13} \ \mathrm{s}$, heat is transferred from bilayer to $10 \ \mathrm{nm}$ thick substrate in 
$10^{-10} \ \mathrm{s}$, heat diffuses transversely to the sample boundaries over 
$10^{-3} \ \mathrm{s}$.\label{fig:heat}}
\end{figure}

\subsection{Scanning micro-diffraction in space and time}

The ability to isolate and compare the response to $\mathrm{fs}$ excitation at
$\mathrm{\mu s}$ and $\mathrm{ms}$ delays is critical to understanding microscopic
heat transport and potentially the formation of meta-stable
phases~\cite{fausti2011light}. The pulse-picking method illustrated in
Fig.~\ref{fig:timing} allows us to perform multi-dimensional, multi-scale
thermometry on $\mathrm{WSe}_2/\mathrm{MoSe}_2$ via the Debye--Waller effect. The detector frame
rate sets the effective repetition rate of the pulse-picked probe, so that the
$\mathrm{kHz}$ frame rate reduces the time required to perform the experiment
from weeks with a conventional CCD to eight hours with the EMPAD. 

Our multi-dimensional, multi-scale thermometry data is shown in  Fig.~\ref{fig:heat}. The three data dimensions are probe delay and probe position (both plotted horizontally) and the relative change in diffraction intensity $\Delta I/I$ (plotted vertically). Colors indicate which spatial cuts in  Fig.~\ref{fig:heat}(b) correspond to the pump--probe delay shown in  Fig.~\ref{fig:heat}(c). Panel (b) i. shows the ultrafast response taken at a pump delay of $20 \ \mathrm{ps}$. The remaining data are taken by progressively extending the duration of the pump pulse train and probing the sample $8 \ \mathrm{\mu s}$ after the final pulse in the pump pulse train. The $8 \ \mathrm{\mu s}$ increment corresponds to the period between pulses at our $125 \ \mathrm{kHz}$ repetition rate. The data in  Fig.~\ref{fig:heat}(b) ii.--v. clearly show the accumulation of energy through the duration of the pump pulse train, while panels vi.--viii. show the dissipation of energy to the sample boundaries after the pump-pulse train terminates.

In the limit that the relaxed sample temperature is well above zero and the temperature change $\Delta T$ is small, both compared to the sample Debye temperature, the fractional change in the Bragg scattering rate $\Delta I/I$ with scattering vector ${\mathbf{k}}$ is proportional to $k^2\Delta T$. Hence, by scanning $\Delta I/I$ as a function of pump spatial position and pump delay, it is possible to map out thermal transport in the sample. We choose a low pump fluence of $2 \ \mathrm{mJ}/\mathrm{cm}^2$ as a compromise between, on the one hand, maintaining linearity in the relationship between $\Delta I/I$ and $T$, and on the other, minimizing Poisson noise in the diffraction difference.

To model the heat transport phenomenology, we fit to our diffraction signal $\Delta I/I \propto T$ to an analytic, approximate solution to the inhomogeneous 2D heat equation,
\begin{equation}
\label{heat}
    \left[\frac{\partial}{\partial t} - \alpha \nabla^2\right] T(t, {\mathbf{x}}; \alpha, A, \sigma_0) = f(t, {\mathbf{x}}; A, \sigma_0).
\end{equation}
The periodic solution $T(t, {\mathbf{x}}; \alpha, A, \sigma_0)$  contains three parameters: the diffusion constant $\alpha$, and the amplitude $A$ and width $\sigma_0$ of the pump pulses. The forcing term $f(t, {\mathbf{x}}; A, \sigma_0)$ represents the pump injecting energy into the system. The model makes no assumptions concerning the transport mechanism. The diffraction intensity $I$ is the sum of peaks labeled 1--4  in  Fig.~\ref{fig:diffraction}(a). Eq.~\eqref{heat} is solvable by Fourier series, and we obtain an analytic approximation by simply truncating this series to finite order. The Supplemental Material provides a detailed derivation of the explicit expression.

We find excellent agreement between our data and three-parameter phenomenological model. Solid lines in  Fig.~\ref{fig:heat}(b) ii--viii show the spatial profile implied by the fitted model; the solid line in  Fig.~\ref{fig:heat}(c) shows the fitted temperature envelope at the center of the sample square. The interpretation of  Fig.~\ref{fig:heat} is that the extreme aspect ratio of the bilayer--substrate combination -- $10~{\rm nm}$ thick versus $250 \ \mathrm{\mu m}$ wide -- results in two relaxation timescales: a fast timescale, $\tau_{\mathrm{fast}} < 1 \ \mathrm{n s}$, and a slow timescale, $ \tau_{\mathrm{slow}} > 1 \ \mathrm{ms}$.  The fit implies a decay in the temperature of the bilayer following the arrival of the first pump pulse to 5\% of its peak value before the arrival of the second. This 5\% residue accumulates for the duration of the $440 \ \mathrm{\mu s}$ pulse train and, following the end of the pulse train, relaxes exponentially at a rate set by the sample window size $L$: $\tau_{\mathrm{slow}} = L^2/(2\pi^2 \alpha)$.

The fit in  Fig.~\ref{fig:heat} gives the value $\tau_{\mathrm{slow}} = 3 \ \mathrm{ms}$. When, as in our experiment, the period between pump pulse trains is less than the relaxation time, the bilayer reaches a periodic state in which the minimum temperature of the pumped region during a cycle remains elevated above the temperature of the boundaries. The estimate of $\tau_{\mathrm{slow}}$  unambiguously defines  the repetition rate that allows for the sample to fully relax before the arrival of each pump pulse in stroboscopic data acquisition. This relaxation time is sample and substrate dependent, as are the physical implications of pumping at a repetition rate faster than sample relaxation. Our method of extending the range of ultrafast pump--probe delays to $\mathrm{\mu s}$--$\mathrm{ms}$ enables us to investigate these issues experimentally.

Pump--probe delays in the $\mathrm{\mu s}$--$\mathrm{ms}$ range provide a technique
to extract the transverse thermal conductivity of the bilayer. This intrinsic
property cannot be inferred from sub-ns data alone, because the ultrafast
relaxation we observe is dominated by the interfacial resistance between
bilayer and SiN substrate, as heat is transferred across the $\mathrm{nm}$
dimension~\cite{britt2022direct}. Whereas, on the slow timescale, the bilayer
and SiN substrate provide parallel channels for conducting heat transversely
over the $10^{-4} \ \mathrm{m}$ distance to the $\mathrm{Si}$ wafer at the transverse
boundary. 

The transverse diffusivity parameter $\alpha$ that we fit with our
phenomenological 2D model (solid lines in  Fig.~\ref{fig:heat}(b)--(c)) does not
discriminate between bilayer and substrate contributions. From the same data,
finite-element simulations (see, e.g.,~\cite{zalden2019femtosecond}) can
extract the intrinsic transverse diffusivity of the bilayer. We find close
agreement between simulation and the approximate analytic relationship,
\begin{equation}
    \alpha_{\mathrm{BL}} = \alpha_{\mathrm{exp}} + \frac{C_{\mathrm{SiN}}}{C_{\mathrm{BL}}}(\alpha_{\mathrm{exp}}-\alpha_{\mathrm{SiN}}),
\end{equation}
where $\alpha_{\mathrm{BL}}$, $C_{\mathrm{BL}}$ are the transverse diffusivity and heat capacity of the bilayer, $\alpha_{\mathrm{SiN}}$, $C_{\mathrm{SiN}}$ are the traverse diffusivity and heat capacity of the substrate, and $\alpha_{\mathrm{exp}}$ is the transverse diffusivity of the bilayer--substrate combination that we measure experimentally.

Hypothetically, data obtained from a freestanding bilayer would be simpler to analyze: however, an implication of the data we collect is that pumping a freestanding bilayer with $2 \ \mathrm{mJ/cm}^{2}$ fluence at $10^2 \ \mathrm{Hz}$ or faster repetition rates would cause irreversible damage within a few pulses, without the SiN present to act as a heatsink.
 In testing samples mounted on SiN, we observed an irreversible drop in
scattering intensity at a fluence of $3 \ \mathrm{mJ/cm}^{2}$. A scanning-micrograph
recorded with our electron probe showed the damage to be localized to the
region illuminated by the pump-laser spot, a conclusion confirmed by
post-mortem optical microscope images of the sample. The ability to reach
higher fluences reversibly is important because  even reversible structural
responses can change discontinuously as a function of
fluence~\cite{sie_ultrafast_2019}, possibly because the out-of-equilibrium
electron population excited by the pump undergoes a phase transition as a
function of the density of excited charges --- in the case of our sample from
an excitonic to a free electron gas~\cite{wang2021diffusivity}. These
fluence-dependent physical mechanisms cannot be fully explored in experiments
where fluence is constrained by the sample environment. 

\section{Discussion and Conclusion}
\label{sec13}

This work has presented new, dramatic advantages of an integrating direct
electron detector with high dynamic range and fast frame rate for structural
dynamics data acquisition. We are able to resolve the ultrafast response of a
$10 \ \mathrm{nm}$ periodic moir\'e superlattice, and to track the $\mathrm{ms}$ long
thermal relaxation of the $\mathrm{WSe}_2/\mathrm{MoSe}_2$ bilayer following ultrafast excitation.
Future experiments plan to apply this technique to investigate the effects of
interlayer interactions on thermal transport in two ways: by measuring the
dependence of transverse relaxation on bilayer twist
angle~\cite{wang2021diffusivity}, and by tuning pump photon energy to
resonantly excite a specific monolayer in the heterostructure.

 Our results show that beam chopping at frequencies up to $500 ~\mathrm{Hz}$
eliminates experimental uncertainties that are caused by non-Poissonian
fluctuations in the probe current on target. In micro-diffraction, an important
non-Poissonian source of uncertainty is positioning error on the probe defining
aperture. Beam chopping is especially important in experiments where detector
saturation or limits on field of view (at fine angular resolution) preclude
measuring the total charge on target per pulse. Beam chopping is a complement to techniques
that eliminate time-of-arrival jitter as a source of experimental
uncertainty~\cite{otto2017solving}.
 
The next generation EMPAD increases the frame rate to
$10 \ \mathrm{kHz}$~\cite{philipp2022very}, which raises the maximum chopping
frequency. Our results suggest that chopping frequencies above $1 \ \mathrm{kHz}$,
while unnecessary for our system in micro-diffraction mode (100 electrons on
target per pulse) 
have the potential to significantly improve signal-to-noise in experiments that involve large bunch charges of $10^5$ electrons or more, and in high-flux x-ray experiments with free electron laser sources.

A natural extension of our pulse-picking technique is to utilize a commercially available femtosecond GHz oscillator and fast pulse picker to achieve  nanosecond pulse selection precision prior to the amplification stage. Such a system, when coupled with a delay stage to cover the range $< 1 \ \mathrm{ns}$, would provide seamless delay capability from femtoseconds to seconds with femtosecond resolution. Measuring the transverse relaxation time of the bilayer, we demonstrate an experimental method for unambiguously defining the sample-dependent optimal repetition-rate for ultrafast stroboscopic data collection. Our results highlight the need for pump--probe modalities that can access multiple time and intensity scales when investigating the rich, multi-scale physics of 2D quantum materials.

This work was supported by the U.S Department of Energy, awards DE-SC0020144 and DE-SC0017631, and U.S. National Science Foundation Grant PHY-1549132, the Center for Bright Beams. D.L., A.S., and A.M.L. acknowledge support from the U.S. Department of Energy,  Office of Science, Basic Energy Sciences, Materials Sciences and Engineering Division, under Contract DE-AC02-76SF00515.

\appendix{}
\section{MUltrafast Electron Diffaction}\label{methods}

\begin{figure}
\includegraphics[width=0.8\columnwidth]{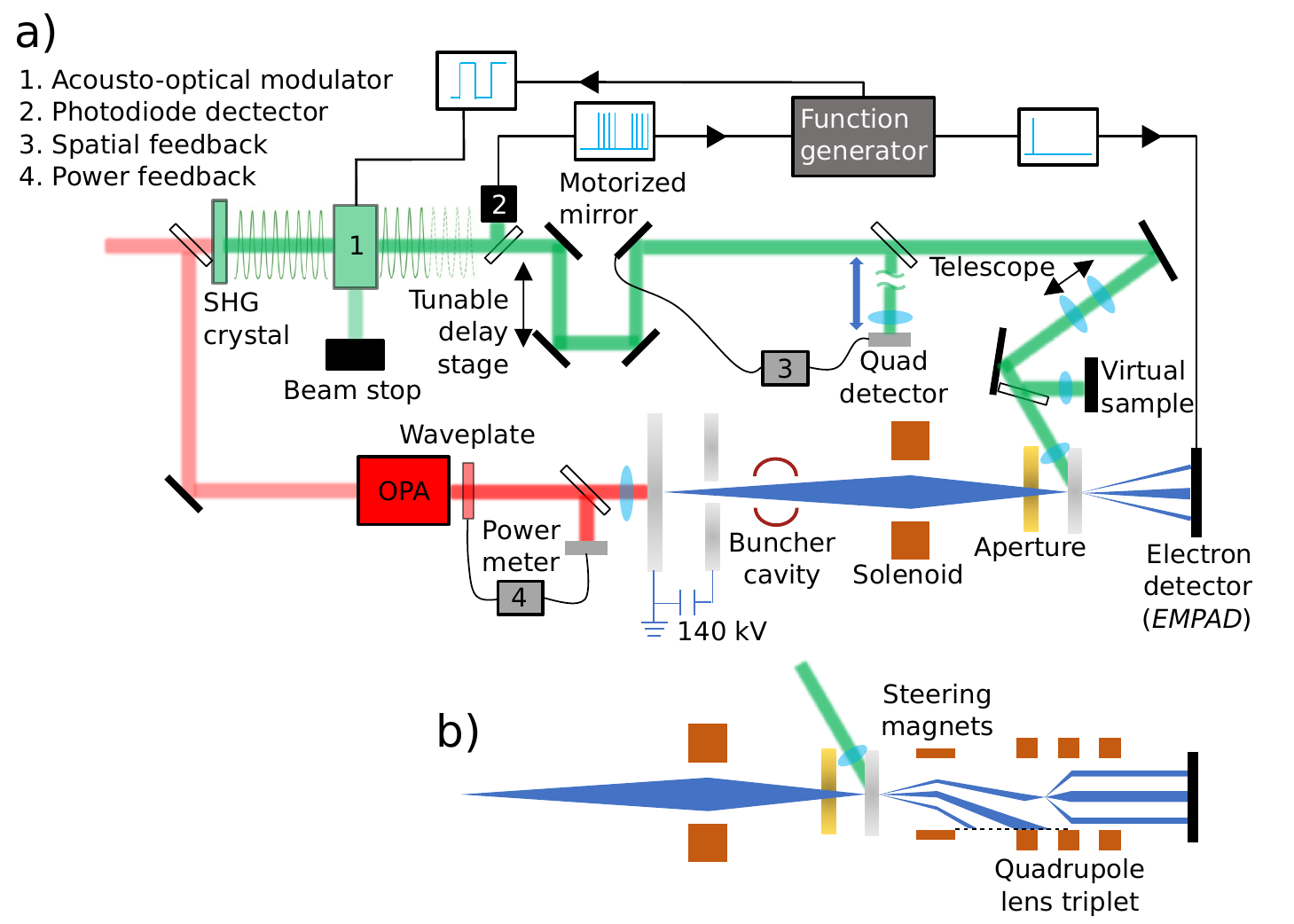}
\caption{(a) Schematic of the UED beamline, see  Ref.~\cite{li2022kiloelectron}
 for details: a $1030 \ \mathrm{nm}$ Yb-fiber laser drives an optical parametric
 amplifier that sends $650 \ \mathrm{nm}$ light pulses to the photocathode.
 Photoemitted bunches are accelerated to $140 \ \mathrm{keV}$, compressed by an
 rf cavity, collimated by a probe-defining aperture, and are collected
 by the detector after scattering on the sample; the same $1030 \ \mathrm{nm}$
 pulses are split and frequency doubled to synchronously pump the
 sample. (b) Modification of the detector section of the beamline to
 accommodate a magnetic quadrupole electron lens triplet. The lens
 triplet enables angular magnification of the scattering pattern. A
 diffraction feature is selected for magnification on the detector with
 a steering magnet upstream of the lens triplet.\label{fig:method}}
\end{figure}

Our beamline is shown in  Fig.~\ref{fig:method}(a). The probe electron beam is photoemitted with 650~nm laser pulses and accelerated to a primary energy of $140 \ \mathrm{keV}$, and the sample is pumped with $515 \ \mathrm{nm}$ pulses obtained from the same $1030 \ \mathrm{nm}$ source. Electron beam optics for magnifying the diffraction pattern are shown in  Fig.~\ref{fig:method}(b). Our scanning ultrafast electron diffraction technique is illustrated in  Fig.~\ref{fig:timing}. The spot size of the electron probe on the sample is defined by a laser-milled aperture $15 \ \mathrm{mm}$ upstream. An in-vacuum lens focuses the $515 \ \mathrm{nm}$ pump pulse to a $10 \ \mathrm{\mu m}$ rms spot on the sample, and the pump spot is steered on the sample by an out-of-vacuum mirror. A virtual-sample camera placed out of vacuum monitors the location of the central peak of the pump laser to $\mathrm{\mu m}$ precision. The duty cycle of the acousto-optic modulator that gates pump pulses is variable to single-pulse precision: we typically choose a duty-cycle to match the detector exposure, eliminating un-detected pulses and thus reducing the thermal load on the sample.  We verify the reliability of the timing system by measuring the total pump-energy per exposure with the detector, and we see a sharp quantization of energy as a function of exposure length at intervals of the $8 \ \mathrm{\mu s}$ laser repetition period. Pump and probe beams are aligned by performing knife-edge scans at the vertical and horizontal sample edges. For the data presented in main text  Fig.~\ref{fig:heat}, knife-edge scans give 6 $\mathrm{\mu m}$ rms probe size in the sample plane.

\section{Detector Saturation}
\label{saturation}

We observe saturation at 60 incident electrons per pixel per pulse in pulsed
operation at $140 \ \mathrm{keV}$. We arrive at this experimental estimate by averaging
single shot measurements, and controlling the charge per pulse by varying the
charge transmitted through the probe-defining aperture. The maximum charge that
the EMPAD can remove per charge-removal cycle per pixel is equivalent to 11
incident electrons at this beam energy. The maximum is set by protection diodes
that limit the peak charging current of each pixel's storage capacitor. There
thus appears to be a discrepancy between, on the one hand, the saturation
threshold we observe and, on the other, the limit set by the electronics. A
plausible resolution of the discrepancy is this: the high charge density in
pulsed operation results in the creation of an electron--hole plasma with
microseconds lifetime. The long-lived plasma is removed over multiple
charge-removal cycles~\cite{weiss2016potential}. The charge-removal circuitry
cycles at $2 \ \mathrm{MHz}$, the probe beam repetition rate is $125 \ \mathrm{kHz}$, and
hence there are 16 charge removal cycles for every probe pulse.

\section{Sample Preparation}

 MoSe$_2$ and WSe$_2$ monolayers are exfoliated from bulk MoSe$_2$ and WSe$_2$ single crystals (HQ graphene)
onto 285~nm $\mathrm{SiO}_2/\mathrm{Si}$ substrate sequentially using a gold tape exfoliation
technique~\cite{liu2020disassembling}, forming
heterostructures with lateral dimensions of mm scale. The crystal orientations of the monolayers in the
heterostructure are aligned with the crystal edges, and further confirmed in electron diffraction. The
heterostructures are later transferred onto 10~nm thick, 250 $\mathrm{\mu m} \times 250 \ \mathrm{\mu m}$ Si$_3$N$_4$ windows on TEM grids
(SiMPore), using a wedging transfer technique with cellulose acetate butyrate
(CAB) polymer~\cite{schneider2010wedging}.

\section{Diffraction simulation}
\label{simulation}

Diffraction patterns in  Fig.~\ref{fig:diffraction}(b) are computed from the
Fourier transform of the real-space phenomenological model presented
in~\cite{zhang2018structural}. Retaining only the longest-wavelength components
of the PLD, atomic positions are displaced by a spatially varying vector field
${\mathbf{u}}({\mathbf{x}})$ with explicit expression,

\numberwithin{equation}{section}
\setcounter{equation}{0}
\begin{align}
    {\mathbf{u}}({\mathbf{x}}) =& \frac{\epsilon}{\Vert  {\mathbf{k}}_0\Vert }\Biggl(\hat{{\mathbf{y}}}\frac{\partial}{\partial x} - \hat{{\mathbf{x}}}\frac{\partial}{\partial y} \Biggr)\Biggl(\cos({\mathbf{k}}_0\cdot {\mathbf{x}}) \notag \\
&+  \cos({\mathbf{k}}_1\cdot {\mathbf{x}}) + \cos([{\mathbf{k}}_0-{\mathbf{k}}_1]\cdot {\mathbf{x}})\Biggr)
\end{align}
where $\epsilon$ is the amplitude parameter and ${\mathbf{k}}_i$ are the reciprocal lattice vectors of the moir\'e superlattice.

\pagebreak

\begin{center}
\textbf{\large Supplemental Material: Multi-scale time-resolved electron diffraction: a case study in moir\'{e} materials}

C. J. R. Duncan, M. Kaemingk, W. H. Li, M. B. Andorf, \\ A. C. Bartnik, A. Galdi, M. Gordon, C. A. Pennington, \\  I. V. Bazarov, H. J. Zeng,  F. Liu, \\ D. Luo, A. Sood, A. M. Lindenberg, M. W. Tate, \\ D. A. Muller, J. Thom-Levy, S. M. Gruner, J. M. Maxson
\end{center}
\setcounter{equation}{0}
\setcounter{figure}{0}
\setcounter{table}{0}
\setcounter{page}{1}
\makeatletter
\renewcommand{\theequation}{S\arabic{equation}}
\renewcommand{\thefigure}{S\arabic{figure}}
\renewcommand{\bibnumfmt}[1]{[S#1]}
\renewcommand{\citenumfont}[1]{S#1}

\textbf{Diffusion Model}

In this supplement we derive an explicit expression for the function used to fit the data in Fig.~5 of the main text.

Our model assumes: i. a square domain having side-length $L=250 \ \mathrm{\mu m}$, ii. that pump pulses strike the center of the square, and iii. that the boundaries of the square are held at constant temperature. We experimentally verify i. With respect to assumption ii., in experiment, we hold the probe spot fixed at the center of the sample window and scan the pump, as shown in Fig.~1 of the main text. Finally, assumption iii. is highly plausible given the overwhelming thermal mass of the Si wafer in which the SiN-supported sample window is embedded.

The fit function $T(x,y,t; \alpha, A, \sigma_0)$ with fit parameters $\alpha, A, \sigma_0$ is a solution to the inhomogenous heat equation:
\begin{equation}\label{PDE}
    \frac{\partial T}{\partial t} - \alpha\left[ \frac{\partial^2 T}{\partial x^2} + \frac{\partial^2 T}{\partial y^2}\right] = f(x,y,t; A, \sigma_0).
\end{equation}
Equation~\eqref{PDE} is subject to the condition that $T\equiv 0$ on a square boundary of side length $L$. The origin of the Cartesian coordinate system $x,y$ lies at the center of the square. To avoid confusion , $\alpha$ is a fit parameter and no assumption is made in the fit as to the mechanism for heat diffusion.

We model the forcing term $f$ in Eq.~\eqref{PDE} as a sequence of delta-function impulses, each having the same Gaussian spatial profile centered at the origin, with amplitude $A$ and r.m.s size $\sigma_0$. Pulses arrive in trains. Trains arrive at the rate $\nu$ and, within each train, pulses arrive at the rate $R$. Each train contains $J$ pulses. The explicit expression for $f$ is then,
\begin{align}
    f(x,y,t; A, \sigma_0) =& \frac{4A}{L^2}\sum_{q=-\infty}^\infty\sum_{j=0}^{J-1}\sum_{m=0}^{M-1}\sum_{n=0}^{N-1}\cos\left((2m+1)\frac{\pi x}{L}\right)\cos\left((2n+1)\frac{\pi y}{L}\right) \exp\big\{\notag\\ 
    &-\frac{\sigma_0^2}{2}\left[(2m+1)^2 + (2n+1)^2 \right]\frac{\pi^2}{L^2} \big\}\delta\left(t-\frac{j}{R}-\frac{q}{\nu}\right).
\end{align}
The sums in $n,m$ are taken over Fourier modes that vanish at the boundary, truncated to orders $M, N = 10$. 

We solve the inhomogeneous problem Eq.~\eqref{PDE} by first solving for the response \newline
$T_0(x,y,t; \alpha, A, \sigma_0)$ to a single forcing pulse, treated as a homogenous problem with the forcing term accounted for in the initial conditions. The linearity of Eq.~\eqref{PDE} then entails that,
\begin{equation}\label{sumthis}
    T(x,y,t;\alpha,A,\sigma_0) = \sum_{q=-\infty}^\infty\sum_{j=0}^{J-1} T_0\left(x,y,t-\frac{j}{R}-\frac{q}{\nu}; \alpha, A, \sigma_0\right).
\end{equation} 
It is well known that for a single spatial Fourier mode $\hat{T}(k,t)$, the solution to the homogenous heat equation is,
\begin{equation}
    \hat{T}(k,t)= \hat{T}(k,0)e^{-\alpha t k^2}.
\end{equation}
For the first impulse we therefore obtain,
\begin{align}\label{temp0}
    T_0(x,y,t; \alpha, A, \sigma_0) = \notag \\ \frac{4A}{L^2}\sum_{m=0}^{M-1}\sum_{n=0}^{N-1}\cos\left((2m+1)\frac{\pi x}{L}\right)\cos\left((2n+1)\frac{\pi y}{L}\right) \exp\big\{\notag\\ 
    -\frac{1}{2}(\sigma_0^2+2\alpha t)\left[(2m+1)^2 + (2n+1)^2 \right]\frac{\pi^2}{L^2} \big\}.
\end{align}

Having in hand the expression for $T_0$ on the right hand side of Eq.~\eqref{temp0}, to compute the sums in Eq.~\eqref{sumthis}, we first approximate the sum inside each train as an integral. It is convenient to define three expressions that appear at intermediate steps in the computation,
\begin{equation}
E_{mn}(t; \alpha) := \exp\left\{-\alpha t \left[(2m+1)^2 + (2n+1)^2\right] \frac{\pi^2}{L^2}  \right\},
\end{equation}
\begin{equation}
   P_{mn}(t; \alpha): =  E_{mn}(t; \alpha)*\sum_{j=0}^{J-1}\delta(t-j/R),
\end{equation}
and,
\begin{equation}
    \tilde{P}_{mn}(t; \alpha):= \sum_{q=-\infty}^\infty P_{mn}(t - q/\nu).
\end{equation}
The approximating integral is then performed piece-wise in time, first for $t < J/R$, inside the pulse train,
\begin{align}
P_{mn}(t; \alpha) =& R \int_0^t E_{mn}(\tau; \alpha) d\tau \\
&= \frac{L^2 R}{\alpha\pi^2\left[(2m+1)^2 + (2n+1)^2\right]}\left[ 1 - E_{mn}(t; \alpha)\right]
\end{align}
Then, after the pulse train has ended, for $t \geq J/R$:
\begin{align}
P_{mn}(t; \alpha) = \frac{L^2 R}{\alpha\pi^2\left[(2m+1)^2 + (2n+1)^2\right]}\left[ 1 - E_{mn}(J/R; \alpha)\right]E_{mn}(t-J/R).
\end{align}

The remaining sum over all pulse trains can be be computed using the formula for a geometric series to give, for $t < J/R$:
\begin{align}
    \tilde{P}_{mn}(t; \alpha) =& \frac{L^2 R}{\alpha\pi^2\left[(2m+1)^2 + (2n+1)^2\right]}\bigg\{ 1 - E_{mn}(t; \alpha) \notag \\ &+ \left[ 1 - E_{mn}(J/R; \alpha)\right]E_{mn}(t-J/R; \alpha)\sum_{j=1}^{\infty} \exp\bigg\{\notag \\&-\frac{1}{2}\alpha j\left[(2m+1)^2 + (2n+1)^2\right] \frac{\pi^2}{fL^2}\bigg\}\bigg\}
    \\
    =&\frac{L^2 R}{\alpha\pi^2\left[(2m+1)^2 + (2n+1)^2\right]}\bigg\{ 1 - E_{mn}(t; \alpha) \notag\\ &+ \left[ 1 - E_{mn}(J/R; \alpha)\right]\frac{E_{mn}(t-J/R; \alpha)}{E_{mn}(-1/\nu; \alpha)-1}\bigg\},
\end{align}
and for $t\geq J/R$,
\begin{align}
    \tilde{P}_{mn}(t; \alpha) \notag \\= \frac{L^2 R}{\alpha\pi^2\left[(2m+1)^2 + (2n+1)^2\right]}\left[ 1 - E_{mn}(J/R; \alpha)\right]\frac{E_{mn}(t-J/R; \alpha)}{1-E_{mn}(1/\nu; \alpha)}.
\end{align}
The fit function expressed in terms of the $\tilde{P}_{mn}$ is therefore,
\begin{align}
    T(x,y,t; \alpha, A, \sigma_0) = \notag \\ \frac{4A}{L^2}\sum_{m=0}^{M-1}\sum_{n=0}^{N-1}\cos\left((2m+1)\frac{\pi x}{L}\right)\cos\left((2n+1)\frac{\pi y}{L}\right)\tilde{P}_{mn}(t; \alpha)\exp\bigg\{ \notag \\
    -\frac{1}{2}\sigma_0^2\left[(2m+1)^2+(2n+1)^2\right]\frac{\pi^2}{L^2}\bigg\}.
\end{align}


\begin{thebibliography}{59}%
\makeatletter
\providecommand \@ifxundefined [1]{%
 \@ifx{#1\undefined}
}%
\providecommand \@ifnum [1]{%
 \ifnum #1\expandafter \@firstoftwo
 \else \expandafter \@secondoftwo
 \fi
}%
\providecommand \@ifx [1]{%
 \ifx #1\expandafter \@firstoftwo
 \else \expandafter \@secondoftwo
 \fi
}%
\providecommand \natexlab [1]{#1}%
\providecommand \enquote  [1]{``#1''}%
\providecommand \bibnamefont  [1]{#1}%
\providecommand \bibfnamefont [1]{#1}%
\providecommand \citenamefont [1]{#1}%
\providecommand \href@noop [0]{\@secondoftwo}%
\providecommand \href [0]{\begingroup \@sanitize@url \@href}%
\providecommand \@href[1]{\@@startlink{#1}\@@href}%
\providecommand \@@href[1]{\endgroup#1\@@endlink}%
\providecommand \@sanitize@url [0]{\catcode `\\12\catcode `\$12\catcode
  `\&12\catcode `\#12\catcode `\^12\catcode `\_12\catcode `\%12\relax}%
\providecommand \@@startlink[1]{}%
\providecommand \@@endlink[0]{}%
\providecommand \url  [0]{\begingroup\@sanitize@url \@url }%
\providecommand \@url [1]{\endgroup\@href {#1}{\urlprefix }}%
\providecommand \urlprefix  [0]{URL }%
\providecommand \Eprint [0]{\href }%
\providecommand \doibase [0]{https://doi.org/}%
\providecommand \selectlanguage [0]{\@gobble}%
\providecommand \bibinfo  [0]{\@secondoftwo}%
\providecommand \bibfield  [0]{\@secondoftwo}%
\providecommand \translation [1]{[#1]}%
\providecommand \BibitemOpen [0]{}%
\providecommand \bibitemStop [0]{}%
\providecommand \bibitemNoStop [0]{.\EOS\space}%
\providecommand \EOS [0]{\spacefactor3000\relax}%
\providecommand \BibitemShut  [1]{\csname bibitem#1\endcsname}%
\let\auto@bib@innerbib\@empty
%</preamble>
\bibitem [{\citenamefont {Ihee}\ \emph {et~al.}(2001)\citenamefont {Ihee},
  \citenamefont {Lobastov}, \citenamefont {Gomez}, \citenamefont {Goodson},
  \citenamefont {Srinivasan}, \citenamefont {Ruan},\ and\ \citenamefont
  {Zewail}}]{ihee2001direct}%
  \BibitemOpen
  \bibfield  {author} {\bibinfo {author} {\bibfnamefont {H.}~\bibnamefont
  {Ihee}}, \bibinfo {author} {\bibfnamefont {V.~A.}\ \bibnamefont {Lobastov}},
  \bibinfo {author} {\bibfnamefont {U.~M.}\ \bibnamefont {Gomez}}, \bibinfo
  {author} {\bibfnamefont {B.~M.}\ \bibnamefont {Goodson}}, \bibinfo {author}
  {\bibfnamefont {R.}~\bibnamefont {Srinivasan}}, \bibinfo {author}
  {\bibfnamefont {C.-Y.}\ \bibnamefont {Ruan}},\ and\ \bibinfo {author}
  {\bibfnamefont {A.~H.}\ \bibnamefont {Zewail}},\ }\bibfield  {title}
  {\bibinfo {title} {Direct imaging of transient molecular structures with
  ultrafast diffraction},\ }\href@noop {} {\bibfield  {journal} {\bibinfo
  {journal} {Science}\ }\textbf {\bibinfo {volume} {291}},\ \bibinfo {pages}
  {458} (\bibinfo {year} {2001})}\BibitemShut {NoStop}%
\bibitem [{\citenamefont {Cavalleri}\ \emph {et~al.}(2001)\citenamefont
  {Cavalleri}, \citenamefont {T{\'o}th}, \citenamefont {Siders}, \citenamefont
  {Squier}, \citenamefont {R{\'a}ksi}, \citenamefont {Forget},\ and\
  \citenamefont {Kieffer}}]{cavalleri2001femtosecond}%
  \BibitemOpen
  \bibfield  {author} {\bibinfo {author} {\bibfnamefont {A.}~\bibnamefont
  {Cavalleri}}, \bibinfo {author} {\bibfnamefont {C.}~\bibnamefont {T{\'o}th}},
  \bibinfo {author} {\bibfnamefont {C.~W.}\ \bibnamefont {Siders}}, \bibinfo
  {author} {\bibfnamefont {J.~A.}\ \bibnamefont {Squier}}, \bibinfo {author}
  {\bibfnamefont {F.}~\bibnamefont {R{\'a}ksi}}, \bibinfo {author}
  {\bibfnamefont {P.}~\bibnamefont {Forget}},\ and\ \bibinfo {author}
  {\bibfnamefont {J.~C.}\ \bibnamefont {Kieffer}},\ }\bibfield  {title}
  {\bibinfo {title} {Femtosecond structural dynamics in {${\mathrm{VO}}_2$}
  during an ultrafast solid-solid phase transition},\ }\href@noop {} {\bibfield
   {journal} {\bibinfo  {journal} {Phys. Rev. Lett.}\ }\textbf {\bibinfo
  {volume} {87}} (\bibinfo {year} {2001})}\BibitemShut {NoStop}%
\bibitem [{\citenamefont {Siwick}\ \emph {et~al.}(2003)\citenamefont {Siwick},
  \citenamefont {Dwyer}, \citenamefont {Jordan},\ and\ \citenamefont
  {Miller}}]{siwick2003atomic}%
  \BibitemOpen
  \bibfield  {author} {\bibinfo {author} {\bibfnamefont {B.~J.}\ \bibnamefont
  {Siwick}}, \bibinfo {author} {\bibfnamefont {J.~R.}\ \bibnamefont {Dwyer}},
  \bibinfo {author} {\bibfnamefont {R.~E.}\ \bibnamefont {Jordan}},\ and\
  \bibinfo {author} {\bibfnamefont {R.~J.~D.}\ \bibnamefont {Miller}},\
  }\bibfield  {title} {\bibinfo {title} {An atomic-level view of melting using
  femtosecond electron diffraction},\ }\href@noop {} {\bibfield  {journal}
  {\bibinfo  {journal} {Science}\ }\textbf {\bibinfo {volume} {302}},\ \bibinfo
  {pages} {1382} (\bibinfo {year} {2003})}\BibitemShut {NoStop}%
\bibitem [{\citenamefont {Gedik}\ \emph {et~al.}(2007)\citenamefont {Gedik},
  \citenamefont {Yang}, \citenamefont {Logvenov}, \citenamefont {Bozovic},\
  and\ \citenamefont {Zewail}}]{gedik2007nonequilibrium}%
  \BibitemOpen
  \bibfield  {author} {\bibinfo {author} {\bibfnamefont {N.}~\bibnamefont
  {Gedik}}, \bibinfo {author} {\bibfnamefont {D.-S.}\ \bibnamefont {Yang}},
  \bibinfo {author} {\bibfnamefont {G.}~\bibnamefont {Logvenov}}, \bibinfo
  {author} {\bibfnamefont {I.}~\bibnamefont {Bozovic}},\ and\ \bibinfo {author}
  {\bibfnamefont {A.~H.}\ \bibnamefont {Zewail}},\ }\bibfield  {title}
  {\bibinfo {title} {Nonequilibrium phase transitions in cuprates observed by
  ultrafast electron crystallography},\ }\href@noop {} {\bibfield  {journal}
  {\bibinfo  {journal} {Science}\ }\textbf {\bibinfo {volume} {316}},\ \bibinfo
  {pages} {425} (\bibinfo {year} {2007})}\BibitemShut {NoStop}%
\bibitem [{\citenamefont {Stojchevska}\ \emph {et~al.}(2014)\citenamefont
  {Stojchevska}, \citenamefont {Vaskivskyi}, \citenamefont {Mertelj},
  \citenamefont {Kusar}, \citenamefont {Svetin}, \citenamefont {Brazovskii},\
  and\ \citenamefont {Mihailovic}}]{stojchevska2014ultrafast}%
  \BibitemOpen
  \bibfield  {author} {\bibinfo {author} {\bibfnamefont {L.}~\bibnamefont
  {Stojchevska}}, \bibinfo {author} {\bibfnamefont {I.}~\bibnamefont
  {Vaskivskyi}}, \bibinfo {author} {\bibfnamefont {T.}~\bibnamefont {Mertelj}},
  \bibinfo {author} {\bibfnamefont {P.}~\bibnamefont {Kusar}}, \bibinfo
  {author} {\bibfnamefont {D.}~\bibnamefont {Svetin}}, \bibinfo {author}
  {\bibfnamefont {S.}~\bibnamefont {Brazovskii}},\ and\ \bibinfo {author}
  {\bibfnamefont {D.}~\bibnamefont {Mihailovic}},\ }\bibfield  {title}
  {\bibinfo {title} {Ultrafast switching to a stable hidden quantum state in an
  electronic crystal},\ }\href@noop {} {\bibfield  {journal} {\bibinfo
  {journal} {Science}\ }\textbf {\bibinfo {volume} {344}},\ \bibinfo {pages}
  {177} (\bibinfo {year} {2014})}\BibitemShut {NoStop}%
\bibitem [{\citenamefont {Sie}\ \emph {et~al.}(2019)\citenamefont {Sie},
  \citenamefont {Nyby}, \citenamefont {Pemmaraju}, \citenamefont {Park},
  \citenamefont {Shen}, \citenamefont {Yang}, \citenamefont {Hoffmann},
  \citenamefont {Ofori-Okai}, \citenamefont {Li}, \citenamefont {Reid},
  \citenamefont {Weathersby}, \citenamefont {Mannebach}, \citenamefont
  {Finney}, \citenamefont {Rhodes}, \citenamefont {Chenet}, \citenamefont
  {Antony}, \citenamefont {Balicas}, \citenamefont {Hone}, \citenamefont
  {Devereaux}, \citenamefont {Heinz}, \citenamefont {Wang},\ and\ \citenamefont
  {Lindenberg}}]{sie_ultrafast_2019}%
  \BibitemOpen
  \bibfield  {author} {\bibinfo {author} {\bibfnamefont {E.~J.}\ \bibnamefont
  {Sie}}, \bibinfo {author} {\bibfnamefont {C.~M.}\ \bibnamefont {Nyby}},
  \bibinfo {author} {\bibfnamefont {C.~D.}\ \bibnamefont {Pemmaraju}}, \bibinfo
  {author} {\bibfnamefont {S.~J.}\ \bibnamefont {Park}}, \bibinfo {author}
  {\bibfnamefont {X.}~\bibnamefont {Shen}}, \bibinfo {author} {\bibfnamefont
  {J.}~\bibnamefont {Yang}}, \bibinfo {author} {\bibfnamefont {M.~C.}\
  \bibnamefont {Hoffmann}}, \bibinfo {author} {\bibfnamefont {B.~K.}\
  \bibnamefont {Ofori-Okai}}, \bibinfo {author} {\bibfnamefont
  {R.}~\bibnamefont {Li}}, \bibinfo {author} {\bibfnamefont {A.~H.}\
  \bibnamefont {Reid}}, \bibinfo {author} {\bibfnamefont {S.}~\bibnamefont
  {Weathersby}}, \bibinfo {author} {\bibfnamefont {E.}~\bibnamefont
  {Mannebach}}, \bibinfo {author} {\bibfnamefont {N.}~\bibnamefont {Finney}},
  \bibinfo {author} {\bibfnamefont {D.}~\bibnamefont {Rhodes}}, \bibinfo
  {author} {\bibfnamefont {D.}~\bibnamefont {Chenet}}, \bibinfo {author}
  {\bibfnamefont {A.}~\bibnamefont {Antony}}, \bibinfo {author} {\bibfnamefont
  {L.}~\bibnamefont {Balicas}}, \bibinfo {author} {\bibfnamefont
  {J.}~\bibnamefont {Hone}}, \bibinfo {author} {\bibfnamefont {T.~P.}\
  \bibnamefont {Devereaux}}, \bibinfo {author} {\bibfnamefont {T.~F.}\
  \bibnamefont {Heinz}}, \bibinfo {author} {\bibfnamefont {X.}~\bibnamefont
  {Wang}},\ and\ \bibinfo {author} {\bibfnamefont {A.~M.}\ \bibnamefont
  {Lindenberg}},\ }\bibfield  {title} {\bibinfo {title} {An ultrafast symmetry
  switch in a {W}eyl semimetal},\ }\href@noop {} {\bibfield  {journal}
  {\bibinfo  {journal} {Nature}\ }\textbf {\bibinfo {volume} {565}},\ \bibinfo
  {pages} {61} (\bibinfo {year} {2019})}\BibitemShut {NoStop}%
\bibitem [{\citenamefont {Broholm}\ \emph {et~al.}(2016)\citenamefont
  {Broholm}, \citenamefont {Fisher}, \citenamefont {Moore}, \citenamefont
  {Murnane}, \citenamefont {Moreo}, \citenamefont {Tranquada}, \citenamefont
  {Basov}, \citenamefont {Freericks}, \citenamefont {Aronson}, \citenamefont
  {MacDonald}, \citenamefont {Fradkin}, \citenamefont {Yacoby}, \citenamefont
  {Samarth}, \citenamefont {Stemmer}, \citenamefont {Horton}, \citenamefont
  {Horwitz}, \citenamefont {Davenport}, \citenamefont {Graf}, \citenamefont
  {Krause}, \citenamefont {Pechan}, \citenamefont {Perry}, \citenamefont
  {Rhyne}, \citenamefont {Schwartz}, \citenamefont {Thiyagarajan},
  \citenamefont {Yarris},\ and\ \citenamefont {Runkles}}]{broholm2016basic}%
  \BibitemOpen
  \bibfield  {author} {\bibinfo {author} {\bibfnamefont {C.}~\bibnamefont
  {Broholm}}, \bibinfo {author} {\bibfnamefont {I.}~\bibnamefont {Fisher}},
  \bibinfo {author} {\bibfnamefont {J.}~\bibnamefont {Moore}}, \bibinfo
  {author} {\bibfnamefont {M.}~\bibnamefont {Murnane}}, \bibinfo {author}
  {\bibfnamefont {A.}~\bibnamefont {Moreo}}, \bibinfo {author} {\bibfnamefont
  {J.}~\bibnamefont {Tranquada}}, \bibinfo {author} {\bibfnamefont
  {D.}~\bibnamefont {Basov}}, \bibinfo {author} {\bibfnamefont
  {J.}~\bibnamefont {Freericks}}, \bibinfo {author} {\bibfnamefont
  {M.}~\bibnamefont {Aronson}}, \bibinfo {author} {\bibfnamefont
  {A.}~\bibnamefont {MacDonald}}, \bibinfo {author} {\bibfnamefont
  {E.}~\bibnamefont {Fradkin}}, \bibinfo {author} {\bibfnamefont
  {A.}~\bibnamefont {Yacoby}}, \bibinfo {author} {\bibfnamefont
  {N.}~\bibnamefont {Samarth}}, \bibinfo {author} {\bibfnamefont
  {S.}~\bibnamefont {Stemmer}}, \bibinfo {author} {\bibfnamefont
  {L.}~\bibnamefont {Horton}}, \bibinfo {author} {\bibfnamefont
  {J.}~\bibnamefont {Horwitz}}, \bibinfo {author} {\bibfnamefont
  {J.}~\bibnamefont {Davenport}}, \bibinfo {author} {\bibfnamefont
  {M.}~\bibnamefont {Graf}}, \bibinfo {author} {\bibfnamefont {J.}~\bibnamefont
  {Krause}}, \bibinfo {author} {\bibfnamefont {M.}~\bibnamefont {Pechan}},
  \bibinfo {author} {\bibfnamefont {K.}~\bibnamefont {Perry}}, \bibinfo
  {author} {\bibfnamefont {J.}~\bibnamefont {Rhyne}}, \bibinfo {author}
  {\bibfnamefont {A.}~\bibnamefont {Schwartz}}, \bibinfo {author}
  {\bibfnamefont {T.}~\bibnamefont {Thiyagarajan}}, \bibinfo {author}
  {\bibfnamefont {L.}~\bibnamefont {Yarris}},\ and\ \bibinfo {author}
  {\bibfnamefont {K.}~\bibnamefont {Runkles}},\ }\bibfield  {title} {\bibinfo
  {title} {Basic research needs workshop on quantum materials for energy
  relevant technology},\ }\href@noop {} {\bibfield  {journal} {\bibinfo
  {journal} {USDOE Off. Sci.}\ } (\bibinfo {year} {2016})}\BibitemShut
  {NoStop}%
\bibitem [{\citenamefont {Carini}\ \emph {et~al.}(2012)\citenamefont {Carini},
  \citenamefont {Denes}, \citenamefont {Gruener},\ and\ \citenamefont
  {Lessner}}]{carini2012neutron}%
  \BibitemOpen
  \bibfield  {author} {\bibinfo {author} {\bibfnamefont {G.}~\bibnamefont
  {Carini}}, \bibinfo {author} {\bibfnamefont {P.}~\bibnamefont {Denes}},
  \bibinfo {author} {\bibfnamefont {S.}~\bibnamefont {Gruener}},\ and\ \bibinfo
  {author} {\bibfnamefont {E.}~\bibnamefont {Lessner}},\ }\bibfield  {title}
  {\bibinfo {title} {Neutron and {X-ray} detectors},\ }\href@noop {} {\bibfield
   {journal} {\bibinfo  {journal} {USDOE Off. Sci.}\ } (\bibinfo {year}
  {2012})}\BibitemShut {NoStop}%
\bibitem [{\citenamefont {Cao}\ \emph {et~al.}(2018)\citenamefont {Cao},
  \citenamefont {Fatemi}, \citenamefont {Fang}, \citenamefont {Watanabe},
  \citenamefont {Taniguchi}, \citenamefont {Kaxiras},\ and\ \citenamefont
  {Jarillo-Herrero}}]{cao2018unconventional}%
  \BibitemOpen
  \bibfield  {author} {\bibinfo {author} {\bibfnamefont {Y.}~\bibnamefont
  {Cao}}, \bibinfo {author} {\bibfnamefont {V.}~\bibnamefont {Fatemi}},
  \bibinfo {author} {\bibfnamefont {S.}~\bibnamefont {Fang}}, \bibinfo {author}
  {\bibfnamefont {K.}~\bibnamefont {Watanabe}}, \bibinfo {author}
  {\bibfnamefont {T.}~\bibnamefont {Taniguchi}}, \bibinfo {author}
  {\bibfnamefont {E.}~\bibnamefont {Kaxiras}},\ and\ \bibinfo {author}
  {\bibfnamefont {P.}~\bibnamefont {Jarillo-Herrero}},\ }\bibfield  {title}
  {\bibinfo {title} {Unconventional superconductivity in magic-angle graphene
  superlattices},\ }\href@noop {} {\bibfield  {journal} {\bibinfo  {journal}
  {Nature}\ }\textbf {\bibinfo {volume} {556}},\ \bibinfo {pages} {43}
  (\bibinfo {year} {2018})}\BibitemShut {NoStop}%
\bibitem [{\citenamefont {Yoo}\ \emph {et~al.}(2019)\citenamefont {Yoo},
  \citenamefont {Engelke}, \citenamefont {Carr}, \citenamefont {Fang},
  \citenamefont {Zhang}, \citenamefont {Cazeaux}, \citenamefont {Sung},
  \citenamefont {Hovden}, \citenamefont {Tsen}, \citenamefont {Taniguchi},
  \citenamefont {Watanabe}, \citenamefont {Yi}, \citenamefont {Kim},
  \citenamefont {Luskin}, \citenamefont {Tadmor}, \citenamefont {Kaxiras},\
  and\ \citenamefont {Kim}}]{yoo2019atomic}%
  \BibitemOpen
  \bibfield  {author} {\bibinfo {author} {\bibfnamefont {H.}~\bibnamefont
  {Yoo}}, \bibinfo {author} {\bibfnamefont {R.}~\bibnamefont {Engelke}},
  \bibinfo {author} {\bibfnamefont {S.}~\bibnamefont {Carr}}, \bibinfo {author}
  {\bibfnamefont {S.}~\bibnamefont {Fang}}, \bibinfo {author} {\bibfnamefont
  {K.}~\bibnamefont {Zhang}}, \bibinfo {author} {\bibfnamefont
  {P.}~\bibnamefont {Cazeaux}}, \bibinfo {author} {\bibfnamefont {S.~H.}\
  \bibnamefont {Sung}}, \bibinfo {author} {\bibfnamefont {R.}~\bibnamefont
  {Hovden}}, \bibinfo {author} {\bibfnamefont {A.~W.}\ \bibnamefont {Tsen}},
  \bibinfo {author} {\bibfnamefont {T.}~\bibnamefont {Taniguchi}}, \bibinfo
  {author} {\bibfnamefont {K.}~\bibnamefont {Watanabe}}, \bibinfo {author}
  {\bibfnamefont {G.-C.}\ \bibnamefont {Yi}}, \bibinfo {author} {\bibfnamefont
  {M.}~\bibnamefont {Kim}}, \bibinfo {author} {\bibfnamefont {M.}~\bibnamefont
  {Luskin}}, \bibinfo {author} {\bibfnamefont {E.~B.}\ \bibnamefont {Tadmor}},
  \bibinfo {author} {\bibfnamefont {E.}~\bibnamefont {Kaxiras}},\ and\ \bibinfo
  {author} {\bibfnamefont {P.}~\bibnamefont {Kim}},\ }\bibfield  {title}
  {\bibinfo {title} {Atomic and electronic reconstruction at the van der
  {Waals} interface in twisted bilayer graphene},\ }\href@noop {} {\bibfield
  {journal} {\bibinfo  {journal} {Nat. Mater.}\ }\textbf {\bibinfo {volume}
  {18}},\ \bibinfo {pages} {448} (\bibinfo {year} {2019})}\BibitemShut
  {NoStop}%
\bibitem [{\citenamefont {Liao}\ \emph {et~al.}(2020)\citenamefont {Liao},
  \citenamefont {Wei}, \citenamefont {Du}, \citenamefont {Wang}, \citenamefont
  {Tang}, \citenamefont {Yu}, \citenamefont {Wu}, \citenamefont {Zhao},
  \citenamefont {Xu}, \citenamefont {Han}, \citenamefont {Liu}, \citenamefont
  {Gao}, \citenamefont {Polcar}, \citenamefont {Sun}, \citenamefont {Shi},
  \citenamefont {Yang},\ and\ \citenamefont {Zhang}}]{liao2020precise}%
  \BibitemOpen
  \bibfield  {author} {\bibinfo {author} {\bibfnamefont {M.}~\bibnamefont
  {Liao}}, \bibinfo {author} {\bibfnamefont {Z.}~\bibnamefont {Wei}}, \bibinfo
  {author} {\bibfnamefont {L.}~\bibnamefont {Du}}, \bibinfo {author}
  {\bibfnamefont {Q.}~\bibnamefont {Wang}}, \bibinfo {author} {\bibfnamefont
  {J.}~\bibnamefont {Tang}}, \bibinfo {author} {\bibfnamefont {H.}~\bibnamefont
  {Yu}}, \bibinfo {author} {\bibfnamefont {F.}~\bibnamefont {Wu}}, \bibinfo
  {author} {\bibfnamefont {J.}~\bibnamefont {Zhao}}, \bibinfo {author}
  {\bibfnamefont {X.}~\bibnamefont {Xu}}, \bibinfo {author} {\bibfnamefont
  {B.}~\bibnamefont {Han}}, \bibinfo {author} {\bibfnamefont {K.}~\bibnamefont
  {Liu}}, \bibinfo {author} {\bibfnamefont {P.}~\bibnamefont {Gao}}, \bibinfo
  {author} {\bibfnamefont {T.}~\bibnamefont {Polcar}}, \bibinfo {author}
  {\bibfnamefont {Z.}~\bibnamefont {Sun}}, \bibinfo {author} {\bibfnamefont
  {D.}~\bibnamefont {Shi}}, \bibinfo {author} {\bibfnamefont {R.}~\bibnamefont
  {Yang}},\ and\ \bibinfo {author} {\bibfnamefont {G.}~\bibnamefont {Zhang}},\
  }\bibfield  {title} {\bibinfo {title} {Precise control of the interlayer
  twist angle in large scale {$\mathrm{MoS}_2$} homostructures},\ }\href@noop
  {} {\bibfield  {journal} {\bibinfo  {journal} {Nat. Commun.}\ }\textbf
  {\bibinfo {volume} {11}},\ \bibinfo {pages} {2153} (\bibinfo {year}
  {2020})}\BibitemShut {NoStop}%
\bibitem [{\citenamefont {Liu}\ \emph {et~al.}(2020)\citenamefont {Liu},
  \citenamefont {Wu}, \citenamefont {Bai}, \citenamefont {Chae}, \citenamefont
  {Li}, \citenamefont {Wang}, \citenamefont {Hone},\ and\ \citenamefont
  {Zhu}}]{liu2020disassembling}%
  \BibitemOpen
  \bibfield  {author} {\bibinfo {author} {\bibfnamefont {F.}~\bibnamefont
  {Liu}}, \bibinfo {author} {\bibfnamefont {W.}~\bibnamefont {Wu}}, \bibinfo
  {author} {\bibfnamefont {Y.}~\bibnamefont {Bai}}, \bibinfo {author}
  {\bibfnamefont {S.~H.}\ \bibnamefont {Chae}}, \bibinfo {author}
  {\bibfnamefont {Q.}~\bibnamefont {Li}}, \bibinfo {author} {\bibfnamefont
  {J.}~\bibnamefont {Wang}}, \bibinfo {author} {\bibfnamefont {J.}~\bibnamefont
  {Hone}},\ and\ \bibinfo {author} {\bibfnamefont {X.-Y.}\ \bibnamefont
  {Zhu}},\ }\bibfield  {title} {\bibinfo {title} {Disassembling { {2D}} van der
  {Waals} crystals into macroscopic monolayers and reassembling into artificial
  lattices},\ }\href@noop {} {\bibfield  {journal} {\bibinfo  {journal}
  {Science}\ }\textbf {\bibinfo {volume} {367}},\ \bibinfo {pages} {903}
  (\bibinfo {year} {2020})}\BibitemShut {NoStop}%
\bibitem [{\citenamefont {Kim}\ \emph {et~al.}(2021)\citenamefont {Kim},
  \citenamefont {Mujid}, \citenamefont {Rai}, \citenamefont {Eriksson},
  \citenamefont {Suh}, \citenamefont {Poddar}, \citenamefont {Ray},
  \citenamefont {Park}, \citenamefont {Fransson}, \citenamefont {Zhong},
  \citenamefont {Muller}, \citenamefont {Erhart}, \citenamefont {Cahill},\ and\
  \citenamefont {Park}}]{kim2021extremely}%
  \BibitemOpen
  \bibfield  {author} {\bibinfo {author} {\bibfnamefont {S.~E.}\ \bibnamefont
  {Kim}}, \bibinfo {author} {\bibfnamefont {F.}~\bibnamefont {Mujid}}, \bibinfo
  {author} {\bibfnamefont {A.}~\bibnamefont {Rai}}, \bibinfo {author}
  {\bibfnamefont {F.}~\bibnamefont {Eriksson}}, \bibinfo {author}
  {\bibfnamefont {J.}~\bibnamefont {Suh}}, \bibinfo {author} {\bibfnamefont
  {P.}~\bibnamefont {Poddar}}, \bibinfo {author} {\bibfnamefont
  {A.}~\bibnamefont {Ray}}, \bibinfo {author} {\bibfnamefont {C.}~\bibnamefont
  {Park}}, \bibinfo {author} {\bibfnamefont {E.}~\bibnamefont {Fransson}},
  \bibinfo {author} {\bibfnamefont {Y.}~\bibnamefont {Zhong}}, \bibinfo
  {author} {\bibfnamefont {D.~A.}\ \bibnamefont {Muller}}, \bibinfo {author}
  {\bibfnamefont {P.}~\bibnamefont {Erhart}}, \bibinfo {author} {\bibfnamefont
  {D.~G.}\ \bibnamefont {Cahill}},\ and\ \bibinfo {author} {\bibfnamefont
  {J.}~\bibnamefont {Park}},\ }\bibfield  {title} {\bibinfo {title} {Extremely
  anisotropic van der {Waals} thermal conductors},\ }\href@noop {} {\bibfield
  {journal} {\bibinfo  {journal} {Nature}\ }\textbf {\bibinfo {volume} {597}},\
  \bibinfo {pages} {660} (\bibinfo {year} {2021})}\BibitemShut {NoStop}%
\bibitem [{\citenamefont {Zhao}\ \emph {et~al.}(2021)\citenamefont {Zhao},
  \citenamefont {Wan}, \citenamefont {Liu}, \citenamefont {Xu}, \citenamefont
  {Yang}, \citenamefont {Shen}, \citenamefont {Zhang}, \citenamefont {Guo},
  \citenamefont {Qian}, \citenamefont {Li}, \citenamefont {Wu}, \citenamefont
  {Lin}, \citenamefont {Yan}, \citenamefont {Li}, \citenamefont {Zhang},
  \citenamefont {Ma}, \citenamefont {Li}, \citenamefont {Chen}, \citenamefont
  {Qiao}, \citenamefont {Shakir}, \citenamefont {Almutairi}, \citenamefont
  {Wei}, \citenamefont {Zhang}, \citenamefont {Pan}, \citenamefont {Huang},
  \citenamefont {Ping}, \citenamefont {Duan},\ and\ \citenamefont
  {Duan}}]{zhao2021high}%
  \BibitemOpen
  \bibfield  {author} {\bibinfo {author} {\bibfnamefont {B.}~\bibnamefont
  {Zhao}}, \bibinfo {author} {\bibfnamefont {Z.}~\bibnamefont {Wan}}, \bibinfo
  {author} {\bibfnamefont {Y.}~\bibnamefont {Liu}}, \bibinfo {author}
  {\bibfnamefont {J.}~\bibnamefont {Xu}}, \bibinfo {author} {\bibfnamefont
  {X.}~\bibnamefont {Yang}}, \bibinfo {author} {\bibfnamefont {D.}~\bibnamefont
  {Shen}}, \bibinfo {author} {\bibfnamefont {Z.}~\bibnamefont {Zhang}},
  \bibinfo {author} {\bibfnamefont {C.}~\bibnamefont {Guo}}, \bibinfo {author}
  {\bibfnamefont {Q.}~\bibnamefont {Qian}}, \bibinfo {author} {\bibfnamefont
  {J.}~\bibnamefont {Li}}, \bibinfo {author} {\bibfnamefont {R.}~\bibnamefont
  {Wu}}, \bibinfo {author} {\bibfnamefont {Z.}~\bibnamefont {Lin}}, \bibinfo
  {author} {\bibfnamefont {X.}~\bibnamefont {Yan}}, \bibinfo {author}
  {\bibfnamefont {B.}~\bibnamefont {Li}}, \bibinfo {author} {\bibfnamefont
  {Z.}~\bibnamefont {Zhang}}, \bibinfo {author} {\bibfnamefont
  {H.}~\bibnamefont {Ma}}, \bibinfo {author} {\bibfnamefont {B.}~\bibnamefont
  {Li}}, \bibinfo {author} {\bibfnamefont {X.}~\bibnamefont {Chen}}, \bibinfo
  {author} {\bibfnamefont {Y.}~\bibnamefont {Qiao}}, \bibinfo {author}
  {\bibfnamefont {I.}~\bibnamefont {Shakir}}, \bibinfo {author} {\bibfnamefont
  {Z.}~\bibnamefont {Almutairi}}, \bibinfo {author} {\bibfnamefont
  {F.}~\bibnamefont {Wei}}, \bibinfo {author} {\bibfnamefont {Y.}~\bibnamefont
  {Zhang}}, \bibinfo {author} {\bibfnamefont {X.}~\bibnamefont {Pan}}, \bibinfo
  {author} {\bibfnamefont {Y.}~\bibnamefont {Huang}}, \bibinfo {author}
  {\bibfnamefont {Y.}~\bibnamefont {Ping}}, \bibinfo {author} {\bibfnamefont
  {X.}~\bibnamefont {Duan}},\ and\ \bibinfo {author} {\bibfnamefont
  {X.}~\bibnamefont {Duan}},\ }\bibfield  {title} {\bibinfo {title} {High-order
  superlattices by rolling up van der {Waals} heterostructures},\ }\href@noop
  {} {\bibfield  {journal} {\bibinfo  {journal} {Nature}\ }\textbf {\bibinfo
  {volume} {591}},\ \bibinfo {pages} {385} (\bibinfo {year}
  {2021})}\BibitemShut {NoStop}%
\bibitem [{\citenamefont {Gadelha}\ \emph {et~al.}(2021)\citenamefont
  {Gadelha}, \citenamefont {Ohlberg}, \citenamefont {Rabelo}, \citenamefont
  {Neto}, \citenamefont {Vasconcelos}, \citenamefont {Campos}, \citenamefont
  {Lemos}, \citenamefont {Ornelas}, \citenamefont {Miranda}, \citenamefont
  {Nadas}, \citenamefont {Santana}, \citenamefont {Watanabe}, \citenamefont
  {Taniguchi}, \citenamefont {van Troeye}, \citenamefont {Lamparski},
  \citenamefont {Meunier}, \citenamefont {Nguyen}, \citenamefont {Paszko},
  \citenamefont {Charlier}, \citenamefont {Campos}, \citenamefont
  {Can{\c{c}}ado}, \citenamefont {Medeiros-Ribeiro},\ and\ \citenamefont
  {Jorio}}]{gadelha2021localization}%
  \BibitemOpen
  \bibfield  {author} {\bibinfo {author} {\bibfnamefont {A.~C.}\ \bibnamefont
  {Gadelha}}, \bibinfo {author} {\bibfnamefont {D.~A.~A.}\ \bibnamefont
  {Ohlberg}}, \bibinfo {author} {\bibfnamefont {C.}~\bibnamefont {Rabelo}},
  \bibinfo {author} {\bibfnamefont {E.~G.~S.}\ \bibnamefont {Neto}}, \bibinfo
  {author} {\bibfnamefont {T.~L.}\ \bibnamefont {Vasconcelos}}, \bibinfo
  {author} {\bibfnamefont {J.~L.}\ \bibnamefont {Campos}}, \bibinfo {author}
  {\bibfnamefont {J.~S.}\ \bibnamefont {Lemos}}, \bibinfo {author}
  {\bibfnamefont {V.}~\bibnamefont {Ornelas}}, \bibinfo {author} {\bibfnamefont
  {D.}~\bibnamefont {Miranda}}, \bibinfo {author} {\bibfnamefont
  {R.}~\bibnamefont {Nadas}}, \bibinfo {author} {\bibfnamefont {F.~C.}\
  \bibnamefont {Santana}}, \bibinfo {author} {\bibfnamefont {K.}~\bibnamefont
  {Watanabe}}, \bibinfo {author} {\bibfnamefont {T.}~\bibnamefont {Taniguchi}},
  \bibinfo {author} {\bibfnamefont {B.}~\bibnamefont {van Troeye}}, \bibinfo
  {author} {\bibfnamefont {M.}~\bibnamefont {Lamparski}}, \bibinfo {author}
  {\bibfnamefont {V.}~\bibnamefont {Meunier}}, \bibinfo {author} {\bibfnamefont
  {V.-H.}\ \bibnamefont {Nguyen}}, \bibinfo {author} {\bibfnamefont
  {D.}~\bibnamefont {Paszko}}, \bibinfo {author} {\bibfnamefont {J.-C.}\
  \bibnamefont {Charlier}}, \bibinfo {author} {\bibfnamefont {L.~C.}\
  \bibnamefont {Campos}}, \bibinfo {author} {\bibfnamefont {L.~G.}\
  \bibnamefont {Can{\c{c}}ado}}, \bibinfo {author} {\bibfnamefont
  {G.}~\bibnamefont {Medeiros-Ribeiro}},\ and\ \bibinfo {author} {\bibfnamefont
  {A.}~\bibnamefont {Jorio}},\ }\bibfield  {title} {\bibinfo {title}
  {Localization of lattice dynamics in low-angle twisted bilayer graphene},\
  }\href@noop {} {\bibfield  {journal} {\bibinfo  {journal} {Nature}\ }\textbf
  {\bibinfo {volume} {590}},\ \bibinfo {pages} {405} (\bibinfo {year}
  {2021})}\BibitemShut {NoStop}%
\bibitem [{\citenamefont {Basov}\ \emph {et~al.}(2017)\citenamefont {Basov},
  \citenamefont {Averitt},\ and\ \citenamefont {Hsieh}}]{basov2017towards}%
  \BibitemOpen
  \bibfield  {author} {\bibinfo {author} {\bibfnamefont {D.}~\bibnamefont
  {Basov}}, \bibinfo {author} {\bibfnamefont {R.}~\bibnamefont {Averitt}},\
  and\ \bibinfo {author} {\bibfnamefont {D.}~\bibnamefont {Hsieh}},\ }\bibfield
   {title} {\bibinfo {title} {Towards properties on demand in quantum
  materials},\ }\href@noop {} {\bibfield  {journal} {\bibinfo  {journal} {Nat.
  Mater.}\ }\textbf {\bibinfo {volume} {16}},\ \bibinfo {pages} {1077}
  (\bibinfo {year} {2017})}\BibitemShut {NoStop}%
\bibitem [{\citenamefont {Andrei}\ \emph {et~al.}(2021)\citenamefont {Andrei},
  \citenamefont {Efetov}, \citenamefont {Jarillo-Herrero}, \citenamefont
  {MacDonald}, \citenamefont {Mak}, \citenamefont {Senthil}, \citenamefont
  {Tutuc}, \citenamefont {Yazdani},\ and\ \citenamefont
  {Young}}]{andrei2021marvels}%
  \BibitemOpen
  \bibfield  {author} {\bibinfo {author} {\bibfnamefont {E.~Y.}\ \bibnamefont
  {Andrei}}, \bibinfo {author} {\bibfnamefont {D.~K.}\ \bibnamefont {Efetov}},
  \bibinfo {author} {\bibfnamefont {P.}~\bibnamefont {Jarillo-Herrero}},
  \bibinfo {author} {\bibfnamefont {A.~H.}\ \bibnamefont {MacDonald}}, \bibinfo
  {author} {\bibfnamefont {K.~F.}\ \bibnamefont {Mak}}, \bibinfo {author}
  {\bibfnamefont {T.}~\bibnamefont {Senthil}}, \bibinfo {author} {\bibfnamefont
  {E.}~\bibnamefont {Tutuc}}, \bibinfo {author} {\bibfnamefont
  {A.}~\bibnamefont {Yazdani}},\ and\ \bibinfo {author} {\bibfnamefont {A.~F.}\
  \bibnamefont {Young}},\ }\bibfield  {title} {\bibinfo {title} {The marvels of
  moir\'e materials},\ }\href@noop {} {\bibfield  {journal} {\bibinfo
  {journal} {Nat. Rev. Mater.}\ }\textbf {\bibinfo {volume} {6}},\ \bibinfo
  {pages} {201} (\bibinfo {year} {2021})}\BibitemShut {NoStop}%
\bibitem [{\citenamefont {Fausti}\ \emph {et~al.}(2011)\citenamefont {Fausti},
  \citenamefont {Tobey}, \citenamefont {Dean}, \citenamefont {Kaiser},
  \citenamefont {Dienst}, \citenamefont {Hoffmann}, \citenamefont {Pyon},
  \citenamefont {Takayama}, \citenamefont {Takagi},\ and\ \citenamefont
  {Cavalleri}}]{fausti2011light}%
  \BibitemOpen
  \bibfield  {author} {\bibinfo {author} {\bibfnamefont {D.}~\bibnamefont
  {Fausti}}, \bibinfo {author} {\bibfnamefont {R.~I.}\ \bibnamefont {Tobey}},
  \bibinfo {author} {\bibfnamefont {N.}~\bibnamefont {Dean}}, \bibinfo {author}
  {\bibfnamefont {S.}~\bibnamefont {Kaiser}}, \bibinfo {author} {\bibfnamefont
  {A.}~\bibnamefont {Dienst}}, \bibinfo {author} {\bibfnamefont {M.~C.}\
  \bibnamefont {Hoffmann}}, \bibinfo {author} {\bibfnamefont {S.}~\bibnamefont
  {Pyon}}, \bibinfo {author} {\bibfnamefont {T.}~\bibnamefont {Takayama}},
  \bibinfo {author} {\bibfnamefont {H.}~\bibnamefont {Takagi}},\ and\ \bibinfo
  {author} {\bibfnamefont {A.}~\bibnamefont {Cavalleri}},\ }\bibfield  {title}
  {\bibinfo {title} {Light-induced superconductivity in a stripe-ordered
  cuprate},\ }\href@noop {} {\bibfield  {journal} {\bibinfo  {journal}
  {Science}\ }\textbf {\bibinfo {volume} {331}},\ \bibinfo {pages} {189}
  (\bibinfo {year} {2011})}\BibitemShut {NoStop}%
\bibitem [{\citenamefont {Weathersby}\ \emph {et~al.}(2015)\citenamefont
  {Weathersby}, \citenamefont {Brown}, \citenamefont {Centurion}, \citenamefont
  {Chase}, \citenamefont {Coffee}, \citenamefont {Corbett}, \citenamefont
  {Eichner}, \citenamefont {Frisch}, \citenamefont {Fry}, \citenamefont
  {G{\"u}hr} \emph {et~al.}}]{weathersby2015mega}%
  \BibitemOpen
  \bibfield  {author} {\bibinfo {author} {\bibfnamefont {S.}~\bibnamefont
  {Weathersby}}, \bibinfo {author} {\bibfnamefont {G.}~\bibnamefont {Brown}},
  \bibinfo {author} {\bibfnamefont {M.}~\bibnamefont {Centurion}}, \bibinfo
  {author} {\bibfnamefont {T.}~\bibnamefont {Chase}}, \bibinfo {author}
  {\bibfnamefont {R.}~\bibnamefont {Coffee}}, \bibinfo {author} {\bibfnamefont
  {J.}~\bibnamefont {Corbett}}, \bibinfo {author} {\bibfnamefont
  {J.}~\bibnamefont {Eichner}}, \bibinfo {author} {\bibfnamefont
  {J.}~\bibnamefont {Frisch}}, \bibinfo {author} {\bibfnamefont
  {A.}~\bibnamefont {Fry}}, \bibinfo {author} {\bibfnamefont {M.}~\bibnamefont
  {G{\"u}hr}}, \emph {et~al.},\ }\bibfield  {title} {\bibinfo {title}
  {Mega-electron-volt ultrafast electron diffraction at { {SLAC}} national
  accelerator laboratory},\ }\href@noop {} {\bibfield  {journal} {\bibinfo
  {journal} {Rev. Sci. Instrum.}\ }\textbf {\bibinfo {volume} {86}} (\bibinfo
  {year} {2015})}\BibitemShut {NoStop}%
\bibitem [{\citenamefont {Feist}\ \emph {et~al.}(2017)\citenamefont {Feist},
  \citenamefont {Bach}, \citenamefont {da~Silva}, \citenamefont {Danz},
  \citenamefont {M{\"o}ller}, \citenamefont {Priebe}, \citenamefont
  {Domr{\"o}se}, \citenamefont {Gatzmann}, \citenamefont {Rost}, \citenamefont
  {Schauss} \emph {et~al.}}]{feist2017ultrafast}%
  \BibitemOpen
  \bibfield  {author} {\bibinfo {author} {\bibfnamefont {A.}~\bibnamefont
  {Feist}}, \bibinfo {author} {\bibfnamefont {N.}~\bibnamefont {Bach}},
  \bibinfo {author} {\bibfnamefont {N.~R.}\ \bibnamefont {da~Silva}}, \bibinfo
  {author} {\bibfnamefont {T.}~\bibnamefont {Danz}}, \bibinfo {author}
  {\bibfnamefont {M.}~\bibnamefont {M{\"o}ller}}, \bibinfo {author}
  {\bibfnamefont {K.~E.}\ \bibnamefont {Priebe}}, \bibinfo {author}
  {\bibfnamefont {T.}~\bibnamefont {Domr{\"o}se}}, \bibinfo {author}
  {\bibfnamefont {J.~G.}\ \bibnamefont {Gatzmann}}, \bibinfo {author}
  {\bibfnamefont {S.}~\bibnamefont {Rost}}, \bibinfo {author} {\bibfnamefont
  {J.}~\bibnamefont {Schauss}}, \emph {et~al.},\ }\bibfield  {title} {\bibinfo
  {title} {Ultrafast transmission electron microscopy using a laser-driven
  field emitter: {F}emtosecond resolution with a high coherence electron
  beam},\ }\href@noop {} {\bibfield  {journal} {\bibinfo  {journal}
  {Ultramicroscopy}\ }\textbf {\bibinfo {volume} {176}},\ \bibinfo {pages} {63}
  (\bibinfo {year} {2017})}\BibitemShut {NoStop}%
\bibitem [{\citenamefont {Li}\ \emph {et~al.}(2020)\citenamefont {Li},
  \citenamefont {Lu}, \citenamefont {Chew}, \citenamefont {Han}, \citenamefont
  {Li}, \citenamefont {Wu}, \citenamefont {Wang}, \citenamefont {Ghimire},\
  and\ \citenamefont {Chang}}]{Li2020HHG}%
  \BibitemOpen
  \bibfield  {author} {\bibinfo {author} {\bibfnamefont {J.}~\bibnamefont
  {Li}}, \bibinfo {author} {\bibfnamefont {J.}~\bibnamefont {Lu}}, \bibinfo
  {author} {\bibfnamefont {A.}~\bibnamefont {Chew}}, \bibinfo {author}
  {\bibfnamefont {S.}~\bibnamefont {Han}}, \bibinfo {author} {\bibfnamefont
  {J.}~\bibnamefont {Li}}, \bibinfo {author} {\bibfnamefont {Y.}~\bibnamefont
  {Wu}}, \bibinfo {author} {\bibfnamefont {H.}~\bibnamefont {Wang}}, \bibinfo
  {author} {\bibfnamefont {S.}~\bibnamefont {Ghimire}},\ and\ \bibinfo {author}
  {\bibfnamefont {Z.}~\bibnamefont {Chang}},\ }\bibfield  {title} {\bibinfo
  {title} {Attosecond science based on high harmonic generation from gases and
  solids},\ }\href@noop {} {\bibfield  {journal} {\bibinfo  {journal} {Nat.
  Commun.}\ }\textbf {\bibinfo {volume} {11}} (\bibinfo {year}
  {2020})}\BibitemShut {NoStop}%
\bibitem [{\citenamefont {Britt}\ \emph {et~al.}(2022)\citenamefont {Britt},
  \citenamefont {Li}, \citenamefont {Ren{\'e}~de Cotret}, \citenamefont
  {Olsen}, \citenamefont {Otto}, \citenamefont {Hassan}, \citenamefont
  {Zacharias}, \citenamefont {Caruso}, \citenamefont {Zhu},\ and\ \citenamefont
  {Siwick}}]{britt2022direct}%
  \BibitemOpen
  \bibfield  {author} {\bibinfo {author} {\bibfnamefont {T.~L.}\ \bibnamefont
  {Britt}}, \bibinfo {author} {\bibfnamefont {Q.}~\bibnamefont {Li}}, \bibinfo
  {author} {\bibfnamefont {L.~P.}\ \bibnamefont {Ren{\'e}~de Cotret}}, \bibinfo
  {author} {\bibfnamefont {N.}~\bibnamefont {Olsen}}, \bibinfo {author}
  {\bibfnamefont {M.}~\bibnamefont {Otto}}, \bibinfo {author} {\bibfnamefont
  {S.~A.}\ \bibnamefont {Hassan}}, \bibinfo {author} {\bibfnamefont
  {M.}~\bibnamefont {Zacharias}}, \bibinfo {author} {\bibfnamefont
  {F.}~\bibnamefont {Caruso}}, \bibinfo {author} {\bibfnamefont
  {X.}~\bibnamefont {Zhu}},\ and\ \bibinfo {author} {\bibfnamefont {B.~J.}\
  \bibnamefont {Siwick}},\ }\bibfield  {title} {\bibinfo {title} {Direct view
  of phonon dynamics in atomically thin {$\mathrm{MoS}_2$}},\ }\href@noop {}
  {\bibfield  {journal} {\bibinfo  {journal} {Nano Lett.}\ } (\bibinfo {year}
  {2022})}\BibitemShut {NoStop}%
\bibitem [{\citenamefont {{R. Meyer}}\ and\ \citenamefont
  {Kirkland}(1998)}]{meyer1998effects}%
  \BibitemOpen
  \bibfield  {author} {\bibinfo {author} {\bibfnamefont {R.}~\bibnamefont {{R.
  Meyer}}}\ and\ \bibinfo {author} {\bibfnamefont {A.}~\bibnamefont
  {Kirkland}},\ }\bibfield  {title} {\bibinfo {title} {The effects of electron
  and photon scattering on signal and noise transfer properties of
  scintillators in { {CCD}} cameras used for electron detection},\ }\href@noop
  {} {\bibfield  {journal} {\bibinfo  {journal} {Ultramicroscopy}\ }\textbf
  {\bibinfo {volume} {75}},\ \bibinfo {pages} {23} (\bibinfo {year}
  {1998})}\BibitemShut {NoStop}%
\bibitem [{\citenamefont {Fan}\ and\ \citenamefont
  {Ellisman}(2000)}]{fan2000digital}%
  \BibitemOpen
  \bibfield  {author} {\bibinfo {author} {\bibfnamefont {G.~Y.}\ \bibnamefont
  {Fan}}\ and\ \bibinfo {author} {\bibfnamefont {M.~H.}\ \bibnamefont
  {Ellisman}},\ }\bibfield  {title} {\bibinfo {title} {Digital imaging in
  transmission electron microscopy},\ }\href@noop {} {\bibfield  {journal}
  {\bibinfo  {journal} {J. Microsc.}\ }\textbf {\bibinfo {volume} {200}},\
  \bibinfo {pages} {1} (\bibinfo {year} {2000})}\BibitemShut {NoStop}%
\bibitem [{\citenamefont {Zuo}(2000)}]{zuo2000electron}%
  \BibitemOpen
  \bibfield  {author} {\bibinfo {author} {\bibfnamefont {J.~M.}\ \bibnamefont
  {Zuo}},\ }\bibfield  {title} {\bibinfo {title} {Electron detection
  characteristics of a slow-scan { {CCD}} camera, imaging plates and film, and
  electron image restoration},\ }\href@noop {} {\bibfield  {journal} {\bibinfo
  {journal} {Microsc. Res. Tech.}\ }\textbf {\bibinfo {volume} {49}},\ \bibinfo
  {pages} {245} (\bibinfo {year} {2000})}\BibitemShut {NoStop}%
\bibitem [{\citenamefont {Gruner}\ \emph {et~al.}(2002)\citenamefont {Gruner},
  \citenamefont {Tate},\ and\ \citenamefont {Eikenberry}}]{gruner2002charge}%
  \BibitemOpen
  \bibfield  {author} {\bibinfo {author} {\bibfnamefont {S.~M.}\ \bibnamefont
  {Gruner}}, \bibinfo {author} {\bibfnamefont {M.~W.}\ \bibnamefont {Tate}},\
  and\ \bibinfo {author} {\bibfnamefont {E.~F.}\ \bibnamefont {Eikenberry}},\
  }\bibfield  {title} {\bibinfo {title} {Charge-coupled device area x-ray
  detectors},\ }\href@noop {} {\bibfield  {journal} {\bibinfo  {journal} {Rev.
  Sci. Instrum.}\ }\textbf {\bibinfo {volume} {73}},\ \bibinfo {pages} {2815}
  (\bibinfo {year} {2002})}\BibitemShut {NoStop}%
\bibitem [{\citenamefont {Ruskin}\ \emph {et~al.}(2013)\citenamefont {Ruskin},
  \citenamefont {Yu},\ and\ \citenamefont
  {Grigorieff}}]{ruskin2013quantitative}%
  \BibitemOpen
  \bibfield  {author} {\bibinfo {author} {\bibfnamefont {R.~S.}\ \bibnamefont
  {Ruskin}}, \bibinfo {author} {\bibfnamefont {Z.}~\bibnamefont {Yu}},\ and\
  \bibinfo {author} {\bibfnamefont {N.}~\bibnamefont {Grigorieff}},\ }\bibfield
   {title} {\bibinfo {title} {Quantitative characterization of electron
  detectors for transmission electron microscopy},\ }\href@noop {} {\bibfield
  {journal} {\bibinfo  {journal} {J. Struct. Biol.}\ }\textbf {\bibinfo
  {volume} {184}},\ \bibinfo {pages} {385} (\bibinfo {year}
  {2013})}\BibitemShut {NoStop}%
\bibitem [{\citenamefont {Vecchione}\ \emph {et~al.}(2017)\citenamefont
  {Vecchione}, \citenamefont {Denes}, \citenamefont {Jobe}, \citenamefont
  {Johnson}, \citenamefont {Joseph}, \citenamefont {Li}, \citenamefont
  {Perazzo}, \citenamefont {Shen}, \citenamefont {Wang}, \citenamefont
  {Weathersby}, \citenamefont {Yang},\ and\ \citenamefont
  {Zhang}}]{vecchione2017direct}%
  \BibitemOpen
  \bibfield  {author} {\bibinfo {author} {\bibfnamefont {T.}~\bibnamefont
  {Vecchione}}, \bibinfo {author} {\bibfnamefont {P.}~\bibnamefont {Denes}},
  \bibinfo {author} {\bibfnamefont {R.~K.}\ \bibnamefont {Jobe}}, \bibinfo
  {author} {\bibfnamefont {I.~J.}\ \bibnamefont {Johnson}}, \bibinfo {author}
  {\bibfnamefont {J.~M.}\ \bibnamefont {Joseph}}, \bibinfo {author}
  {\bibfnamefont {R.~K.}\ \bibnamefont {Li}}, \bibinfo {author} {\bibfnamefont
  {A.}~\bibnamefont {Perazzo}}, \bibinfo {author} {\bibfnamefont
  {X.}~\bibnamefont {Shen}}, \bibinfo {author} {\bibfnamefont {X.~J.}\
  \bibnamefont {Wang}}, \bibinfo {author} {\bibfnamefont {S.~P.}\ \bibnamefont
  {Weathersby}}, \bibinfo {author} {\bibfnamefont {J.}~\bibnamefont {Yang}},\
  and\ \bibinfo {author} {\bibfnamefont {D.}~\bibnamefont {Zhang}},\ }\bibfield
   {title} {\bibinfo {title} {A direct electron detector for time-resolved {
  {MeV}} electron microscopy},\ }\href@noop {} {\bibfield  {journal} {\bibinfo
  {journal} {Rev. Sci. Instrum.}\ }\textbf {\bibinfo {volume} {88}} (\bibinfo
  {year} {2017})}\BibitemShut {NoStop}%
\bibitem [{\citenamefont {Allahgholi}\ \emph {et~al.}(2019)\citenamefont
  {Allahgholi}, \citenamefont {Becker}, \citenamefont {Delfs}, \citenamefont
  {Dinapoli}, \citenamefont {G{\"o}ttlicher}, \citenamefont {Graafsma},
  \citenamefont {Greiffenberg}, \citenamefont {Hirsemann}, \citenamefont
  {Jack}, \citenamefont {Klyuev}, \citenamefont {Kr{\"u}ger}, \citenamefont
  {Kuhn}, \citenamefont {Laurus}, \citenamefont {Marras}, \citenamefont
  {Mezza}, \citenamefont {Mozzanica}, \citenamefont {Poehlsen}, \citenamefont
  {{Shefer Shalev}}, \citenamefont {Sheviakov}, \citenamefont {Schmitt},
  \citenamefont {Schwandt}, \citenamefont {Shi}, \citenamefont {Smoljanin},
  \citenamefont {Trunk}, \citenamefont {Zhang},\ and\ \citenamefont
  {Zimmer}}]{allahgholi2019megapixels}%
  \BibitemOpen
  \bibfield  {author} {\bibinfo {author} {\bibfnamefont {A.}~\bibnamefont
  {Allahgholi}}, \bibinfo {author} {\bibfnamefont {J.}~\bibnamefont {Becker}},
  \bibinfo {author} {\bibfnamefont {A.}~\bibnamefont {Delfs}}, \bibinfo
  {author} {\bibfnamefont {R.}~\bibnamefont {Dinapoli}}, \bibinfo {author}
  {\bibfnamefont {P.}~\bibnamefont {G{\"o}ttlicher}}, \bibinfo {author}
  {\bibfnamefont {H.}~\bibnamefont {Graafsma}}, \bibinfo {author}
  {\bibfnamefont {D.}~\bibnamefont {Greiffenberg}}, \bibinfo {author}
  {\bibfnamefont {H.}~\bibnamefont {Hirsemann}}, \bibinfo {author}
  {\bibfnamefont {S.}~\bibnamefont {Jack}}, \bibinfo {author} {\bibfnamefont
  {A.}~\bibnamefont {Klyuev}}, \bibinfo {author} {\bibfnamefont
  {H.}~\bibnamefont {Kr{\"u}ger}}, \bibinfo {author} {\bibfnamefont
  {M.}~\bibnamefont {Kuhn}}, \bibinfo {author} {\bibfnamefont {T.}~\bibnamefont
  {Laurus}}, \bibinfo {author} {\bibfnamefont {A.}~\bibnamefont {Marras}},
  \bibinfo {author} {\bibfnamefont {D.}~\bibnamefont {Mezza}}, \bibinfo
  {author} {\bibfnamefont {A.}~\bibnamefont {Mozzanica}}, \bibinfo {author}
  {\bibfnamefont {J.}~\bibnamefont {Poehlsen}}, \bibinfo {author}
  {\bibfnamefont {O.}~\bibnamefont {{Shefer Shalev}}}, \bibinfo {author}
  {\bibfnamefont {I.}~\bibnamefont {Sheviakov}}, \bibinfo {author}
  {\bibfnamefont {B.}~\bibnamefont {Schmitt}}, \bibinfo {author} {\bibfnamefont
  {J.}~\bibnamefont {Schwandt}}, \bibinfo {author} {\bibfnamefont
  {X.}~\bibnamefont {Shi}}, \bibinfo {author} {\bibfnamefont {S.}~\bibnamefont
  {Smoljanin}}, \bibinfo {author} {\bibfnamefont {U.}~\bibnamefont {Trunk}},
  \bibinfo {author} {\bibfnamefont {J.}~\bibnamefont {Zhang}},\ and\ \bibinfo
  {author} {\bibfnamefont {M.}~\bibnamefont {Zimmer}},\ }\bibfield  {title}
  {\bibinfo {title} {Megapixels @ {Megahertz -- The} { {AGIPD}} high-speed
  cameras for the { {European} {XFEL}}},\ }\href@noop {} {\bibfield  {journal}
  {\bibinfo  {journal} {Nucl. Instrum.}\ }\textbf {\bibinfo {volume} {942}}
  (\bibinfo {year} {2019})}\BibitemShut {NoStop}%
\bibitem [{\citenamefont {Leonarski}\ \emph {et~al.}(2018)\citenamefont
  {Leonarski}, \citenamefont {Redford}, \citenamefont {Mozzanica},
  \citenamefont {Lopez-Cuenca}, \citenamefont {Panepucci}, \citenamefont
  {Nass}, \citenamefont {Ozerov}, \citenamefont {Vera}, \citenamefont
  {Olieric}, \citenamefont {Buntschu}, \citenamefont {Schneider}, \citenamefont
  {Tinti}, \citenamefont {Froejdh}, \citenamefont {Diederichs}, \citenamefont
  {Bunk}, \citenamefont {Schmitt},\ and\ \citenamefont
  {Wang}}]{leonarski2018fast}%
  \BibitemOpen
  \bibfield  {author} {\bibinfo {author} {\bibfnamefont {F.}~\bibnamefont
  {Leonarski}}, \bibinfo {author} {\bibfnamefont {S.}~\bibnamefont {Redford}},
  \bibinfo {author} {\bibfnamefont {A.}~\bibnamefont {Mozzanica}}, \bibinfo
  {author} {\bibfnamefont {C.}~\bibnamefont {Lopez-Cuenca}}, \bibinfo {author}
  {\bibfnamefont {E.}~\bibnamefont {Panepucci}}, \bibinfo {author}
  {\bibfnamefont {K.}~\bibnamefont {Nass}}, \bibinfo {author} {\bibfnamefont
  {D.}~\bibnamefont {Ozerov}}, \bibinfo {author} {\bibfnamefont
  {L.}~\bibnamefont {Vera}}, \bibinfo {author} {\bibfnamefont {V.}~\bibnamefont
  {Olieric}}, \bibinfo {author} {\bibfnamefont {D.}~\bibnamefont {Buntschu}},
  \bibinfo {author} {\bibfnamefont {R.}~\bibnamefont {Schneider}}, \bibinfo
  {author} {\bibfnamefont {G.}~\bibnamefont {Tinti}}, \bibinfo {author}
  {\bibfnamefont {E.}~\bibnamefont {Froejdh}}, \bibinfo {author} {\bibfnamefont
  {K.}~\bibnamefont {Diederichs}}, \bibinfo {author} {\bibfnamefont
  {O.}~\bibnamefont {Bunk}}, \bibinfo {author} {\bibfnamefont {B.}~\bibnamefont
  {Schmitt}},\ and\ \bibinfo {author} {\bibfnamefont {M.}~\bibnamefont
  {Wang}},\ }\bibfield  {title} {\bibinfo {title} {Fast and accurate data
  collection for macromolecular crystallography using the jungfrau detector},\
  }\href@noop {} {\bibfield  {journal} {\bibinfo  {journal} {Nat. Mater.et}\
  }\textbf {\bibinfo {volume} {15}},\ \bibinfo {pages} {799} (\bibinfo {year}
  {2018})}\BibitemShut {NoStop}%
\bibitem [{\citenamefont {Tate}\ \emph {et~al.}(2013)\citenamefont {Tate},
  \citenamefont {Chamberlain}, \citenamefont {Green}, \citenamefont {Philipp},
  \citenamefont {Purohit}, \citenamefont {Strohman},\ and\ \citenamefont
  {Gruner}}]{tate2013medium}%
  \BibitemOpen
  \bibfield  {author} {\bibinfo {author} {\bibfnamefont {M.~W.}\ \bibnamefont
  {Tate}}, \bibinfo {author} {\bibfnamefont {D.}~\bibnamefont {Chamberlain}},
  \bibinfo {author} {\bibfnamefont {K.~S.}\ \bibnamefont {Green}}, \bibinfo
  {author} {\bibfnamefont {H.~T.}\ \bibnamefont {Philipp}}, \bibinfo {author}
  {\bibfnamefont {P.}~\bibnamefont {Purohit}}, \bibinfo {author} {\bibfnamefont
  {C.}~\bibnamefont {Strohman}},\ and\ \bibinfo {author} {\bibfnamefont
  {S.~M.}\ \bibnamefont {Gruner}},\ }\bibfield  {title} {\bibinfo {title} {A
  medium-format, mixed-mode pixel array detector for {Kilohertz} {X-ray}
  imaging},\ }\href@noop {} {\bibfield  {journal} {\bibinfo  {journal} {J.
  Phys. Conf. Ser.}\ }\textbf {\bibinfo {volume} {425}} (\bibinfo {year}
  {2013})}\BibitemShut {NoStop}%
\bibitem [{\citenamefont {Jiang}\ \emph {et~al.}(2018)\citenamefont {Jiang},
  \citenamefont {Chen}, \citenamefont {Han}, \citenamefont {Deb}, \citenamefont
  {Gao}, \citenamefont {Xie}, \citenamefont {Purohit}, \citenamefont {Tate},
  \citenamefont {Park}, \citenamefont {Gruner}, \citenamefont {Elser},\ and\
  \citenamefont {Muller}}]{jiang2018electron}%
  \BibitemOpen
  \bibfield  {author} {\bibinfo {author} {\bibfnamefont {Y.}~\bibnamefont
  {Jiang}}, \bibinfo {author} {\bibfnamefont {Z.}~\bibnamefont {Chen}},
  \bibinfo {author} {\bibfnamefont {Y.}~\bibnamefont {Han}}, \bibinfo {author}
  {\bibfnamefont {P.}~\bibnamefont {Deb}}, \bibinfo {author} {\bibfnamefont
  {H.}~\bibnamefont {Gao}}, \bibinfo {author} {\bibfnamefont {S.}~\bibnamefont
  {Xie}}, \bibinfo {author} {\bibfnamefont {P.}~\bibnamefont {Purohit}},
  \bibinfo {author} {\bibfnamefont {M.~W.}\ \bibnamefont {Tate}}, \bibinfo
  {author} {\bibfnamefont {J.}~\bibnamefont {Park}}, \bibinfo {author}
  {\bibfnamefont {S.~M.}\ \bibnamefont {Gruner}}, \bibinfo {author}
  {\bibfnamefont {V.}~\bibnamefont {Elser}},\ and\ \bibinfo {author}
  {\bibfnamefont {D.~A.}\ \bibnamefont {Muller}},\ }\bibfield  {title}
  {\bibinfo {title} {Electron ptychography of {{ {2D}}} materials to deep
  sub-{\r{a}}ngstr\"om resolution},\ }\href@noop {} {\bibfield  {journal}
  {\bibinfo  {journal} {Nature}\ }\textbf {\bibinfo {volume} {559}},\ \bibinfo
  {pages} {343} (\bibinfo {year} {2018})}\BibitemShut {NoStop}%
\bibitem [{\citenamefont {Chen}\ \emph {et~al.}(2021)\citenamefont {Chen},
  \citenamefont {Jiang}, \citenamefont {Shao}, \citenamefont {Holtz},
  \citenamefont {Odstril}, \citenamefont {Guizar-Sicairos}, \citenamefont
  {Hanke}, \citenamefont {Ganschow}, \citenamefont {Schlom},\ and\
  \citenamefont {Muller}}]{chen2021electron}%
  \BibitemOpen
  \bibfield  {author} {\bibinfo {author} {\bibfnamefont {Z.}~\bibnamefont
  {Chen}}, \bibinfo {author} {\bibfnamefont {Y.}~\bibnamefont {Jiang}},
  \bibinfo {author} {\bibfnamefont {Y.-T.}\ \bibnamefont {Shao}}, \bibinfo
  {author} {\bibfnamefont {M.~E.}\ \bibnamefont {Holtz}}, \bibinfo {author}
  {\bibfnamefont {M.}~\bibnamefont {Odstril}}, \bibinfo {author} {\bibfnamefont
  {M.}~\bibnamefont {Guizar-Sicairos}}, \bibinfo {author} {\bibfnamefont
  {I.}~\bibnamefont {Hanke}}, \bibinfo {author} {\bibfnamefont
  {S.}~\bibnamefont {Ganschow}}, \bibinfo {author} {\bibfnamefont {D.~G.}\
  \bibnamefont {Schlom}},\ and\ \bibinfo {author} {\bibfnamefont {D.~A.}\
  \bibnamefont {Muller}},\ }\bibfield  {title} {\bibinfo {title} {Electron
  ptychography achieves atomic-resolution limits set by lattice vibrations},\
  }\href@noop {} {\bibfield  {journal} {\bibinfo  {journal} {Science}\ }\textbf
  {\bibinfo {volume} {372}},\ \bibinfo {pages} {826} (\bibinfo {year}
  {2021})}\BibitemShut {NoStop}%
\bibitem [{\citenamefont {Hart}\ \emph {et~al.}(2017)\citenamefont {Hart},
  \citenamefont {Lang}, \citenamefont {Leff}, \citenamefont {Longo},
  \citenamefont {Trevor}, \citenamefont {Twesten},\ and\ \citenamefont
  {Taheri}}]{hart2017}%
  \BibitemOpen
  \bibfield  {author} {\bibinfo {author} {\bibfnamefont {J.~L.}\ \bibnamefont
  {Hart}}, \bibinfo {author} {\bibfnamefont {A.~C.}\ \bibnamefont {Lang}},
  \bibinfo {author} {\bibfnamefont {A.~C.}\ \bibnamefont {Leff}}, \bibinfo
  {author} {\bibfnamefont {P.}~\bibnamefont {Longo}}, \bibinfo {author}
  {\bibfnamefont {C.}~\bibnamefont {Trevor}}, \bibinfo {author} {\bibfnamefont
  {R.~D.}\ \bibnamefont {Twesten}},\ and\ \bibinfo {author} {\bibfnamefont
  {M.~L.}\ \bibnamefont {Taheri}},\ }\bibfield  {title} {\bibinfo {title}
  {Direct detection electron energy-loss spectroscopy: a method to push the
  limits of resolution and sensitivity},\ }\href@noop {} {\bibfield  {journal}
  {\bibinfo  {journal} {Sci. Rep.}\ }\textbf {\bibinfo {volume} {7}},\ \bibinfo
  {pages} {1} (\bibinfo {year} {2017})}\BibitemShut {NoStop}%
\bibitem [{\citenamefont {Shen}(2018)}]{shen2018}%
  \BibitemOpen
  \bibfield  {author} {\bibinfo {author} {\bibfnamefont {P.~S.}\ \bibnamefont
  {Shen}},\ }\bibfield  {title} {\bibinfo {title} {The 2017 {Nobel Prize in
  Chemistry}: cryo-{ {EM}} comes of age},\ }\href@noop {} {\bibfield  {journal}
  {\bibinfo  {journal} {Anal. Bioanal. Chem.}\ }\textbf {\bibinfo {volume}
  {410}},\ \bibinfo {pages} {2053} (\bibinfo {year} {2018})}\BibitemShut
  {NoStop}%
\bibitem [{\citenamefont {Faruqi}\ and\ \citenamefont
  {Henderson}(2007)}]{faruqi2007electronic}%
  \BibitemOpen
  \bibfield  {author} {\bibinfo {author} {\bibfnamefont {A.}~\bibnamefont
  {Faruqi}}\ and\ \bibinfo {author} {\bibfnamefont {R.}~\bibnamefont
  {Henderson}},\ }\bibfield  {title} {\bibinfo {title} {Electronic detectors
  for electron microscopy},\ }\href@noop {} {\bibfield  {journal} {\bibinfo
  {journal} {Curr. Opin. Struct. Biol.}\ }\textbf {\bibinfo {volume} {17}},\
  \bibinfo {pages} {549} (\bibinfo {year} {2007})}\BibitemShut {NoStop}%
\bibitem [{\citenamefont {Lee}\ \emph {et~al.}(2017)\citenamefont {Lee},
  \citenamefont {Kim}, \citenamefont {Kim},\ and\ \citenamefont
  {Kwon}}]{lee2017ultrafast}%
  \BibitemOpen
  \bibfield  {author} {\bibinfo {author} {\bibfnamefont {Y.~M.}\ \bibnamefont
  {Lee}}, \bibinfo {author} {\bibfnamefont {Y.~J.}\ \bibnamefont {Kim}},
  \bibinfo {author} {\bibfnamefont {Y.-J.}\ \bibnamefont {Kim}},\ and\ \bibinfo
  {author} {\bibfnamefont {O.-H.}\ \bibnamefont {Kwon}},\ }\bibfield  {title}
  {\bibinfo {title} {Ultrafast electron microscopy integrated with a direct
  electron detection camera},\ }\href@noop {} {\bibfield  {journal} {\bibinfo
  {journal} {Struct. Dyn.}\ }\textbf {\bibinfo {volume} {4}} (\bibinfo {year}
  {2017})}\BibitemShut {NoStop}%
\bibitem [{\citenamefont {Tate}\ \emph {et~al.}(2016)\citenamefont {Tate},
  \citenamefont {Purohit}, \citenamefont {Chamberlain}, \citenamefont {Nguyen},
  \citenamefont {Hovden}, \citenamefont {Chang}, \citenamefont {Deb},
  \citenamefont {Turgut}, \citenamefont {Heron}, \citenamefont {Schlom} \emph
  {et~al.}}]{tate2016high}%
  \BibitemOpen
  \bibfield  {author} {\bibinfo {author} {\bibfnamefont {M.~W.}\ \bibnamefont
  {Tate}}, \bibinfo {author} {\bibfnamefont {P.}~\bibnamefont {Purohit}},
  \bibinfo {author} {\bibfnamefont {D.}~\bibnamefont {Chamberlain}}, \bibinfo
  {author} {\bibfnamefont {K.~X.}\ \bibnamefont {Nguyen}}, \bibinfo {author}
  {\bibfnamefont {R.}~\bibnamefont {Hovden}}, \bibinfo {author} {\bibfnamefont
  {C.~S.}\ \bibnamefont {Chang}}, \bibinfo {author} {\bibfnamefont
  {P.}~\bibnamefont {Deb}}, \bibinfo {author} {\bibfnamefont {E.}~\bibnamefont
  {Turgut}}, \bibinfo {author} {\bibfnamefont {J.~T.}\ \bibnamefont {Heron}},
  \bibinfo {author} {\bibfnamefont {D.~G.}\ \bibnamefont {Schlom}}, \emph
  {et~al.},\ }\bibfield  {title} {\bibinfo {title} {High dynamic range pixel
  array detector for scanning transmission electron microscopy},\ }\href@noop
  {} {\bibfield  {journal} {\bibinfo  {journal} {Microsc. Microanal.}\ }\textbf
  {\bibinfo {volume} {22}},\ \bibinfo {pages} {237} (\bibinfo {year}
  {2016})}\BibitemShut {NoStop}%
\bibitem [{\citenamefont {Li}\ \emph {et~al.}(2022)\citenamefont {Li},
  \citenamefont {Duncan}, \citenamefont {Andorf}, \citenamefont {Bartnik},
  \citenamefont {Bianco}, \citenamefont {Cultrera}, \citenamefont {Galdi},
  \citenamefont {Gordon}, \citenamefont {Kaemingk}, \citenamefont {Pennington},
  \citenamefont {Kourkoutis}, \citenamefont {Bazarov},\ and\ \citenamefont
  {Maxson}}]{li2022kiloelectron}%
  \BibitemOpen
  \bibfield  {author} {\bibinfo {author} {\bibfnamefont {W.~H.}\ \bibnamefont
  {Li}}, \bibinfo {author} {\bibfnamefont {C.~J.~R.}\ \bibnamefont {Duncan}},
  \bibinfo {author} {\bibfnamefont {M.~B.}\ \bibnamefont {Andorf}}, \bibinfo
  {author} {\bibfnamefont {A.~C.}\ \bibnamefont {Bartnik}}, \bibinfo {author}
  {\bibfnamefont {E.}~\bibnamefont {Bianco}}, \bibinfo {author} {\bibfnamefont
  {L.}~\bibnamefont {Cultrera}}, \bibinfo {author} {\bibfnamefont
  {A.}~\bibnamefont {Galdi}}, \bibinfo {author} {\bibfnamefont
  {M.}~\bibnamefont {Gordon}}, \bibinfo {author} {\bibfnamefont
  {M.}~\bibnamefont {Kaemingk}}, \bibinfo {author} {\bibfnamefont {C.~A.}\
  \bibnamefont {Pennington}}, \bibinfo {author} {\bibfnamefont {L.~F.}\
  \bibnamefont {Kourkoutis}}, \bibinfo {author} {\bibfnamefont {I.~V.}\
  \bibnamefont {Bazarov}},\ and\ \bibinfo {author} {\bibfnamefont {J.~M.}\
  \bibnamefont {Maxson}},\ }\bibfield  {title} {\bibinfo {title} {A
  kiloelectron-volt ultrafast electron micro-diffraction apparatus using low
  emittance semiconductor photocathodes},\ }\href@noop {} {\bibfield  {journal}
  {\bibinfo  {journal} {Struct. Dyn.}\ }\textbf {\bibinfo {volume} {9}}
  (\bibinfo {year} {2022})}\BibitemShut {NoStop}%
\bibitem [{\citenamefont {Shen}\ \emph {et~al.}(2018)\citenamefont {Shen},
  \citenamefont {Li}, \citenamefont {Lundstr{\"o}m}, \citenamefont {Lane},
  \citenamefont {Reid}, \citenamefont {Weathersby},\ and\ \citenamefont
  {Wang}}]{shen2018femtosecond}%
  \BibitemOpen
  \bibfield  {author} {\bibinfo {author} {\bibfnamefont {X.}~\bibnamefont
  {Shen}}, \bibinfo {author} {\bibfnamefont {R.}~\bibnamefont {Li}}, \bibinfo
  {author} {\bibfnamefont {U.}~\bibnamefont {Lundstr{\"o}m}}, \bibinfo {author}
  {\bibfnamefont {T.}~\bibnamefont {Lane}}, \bibinfo {author} {\bibfnamefont
  {A.}~\bibnamefont {Reid}}, \bibinfo {author} {\bibfnamefont {S.}~\bibnamefont
  {Weathersby}},\ and\ \bibinfo {author} {\bibfnamefont {X.}~\bibnamefont
  {Wang}},\ }\bibfield  {title} {\bibinfo {title} {Femtosecond
  mega-electron-volt electron microdiffraction},\ }\href@noop {} {\bibfield
  {journal} {\bibinfo  {journal} {Ultramicroscopy}\ }\textbf {\bibinfo {volume}
  {184}},\ \bibinfo {pages} {172} (\bibinfo {year} {2018})}\BibitemShut
  {NoStop}%
\bibitem [{\citenamefont {Schriever}\ \emph {et~al.}(2008)\citenamefont
  {Schriever}, \citenamefont {Lochbrunner}, \citenamefont {Riedle},\ and\
  \citenamefont {Nesbitt}}]{schriever2008ultrasensitive}%
  \BibitemOpen
  \bibfield  {author} {\bibinfo {author} {\bibfnamefont {C.}~\bibnamefont
  {Schriever}}, \bibinfo {author} {\bibfnamefont {S.}~\bibnamefont
  {Lochbrunner}}, \bibinfo {author} {\bibfnamefont {E.}~\bibnamefont
  {Riedle}},\ and\ \bibinfo {author} {\bibfnamefont {D.}~\bibnamefont
  {Nesbitt}},\ }\bibfield  {title} {\bibinfo {title} {Ultrasensitive
  ultraviolet-visible 20 fs absorption spectroscopy of low vapor pressure
  molecules in the gas phase},\ }\href@noop {} {\bibfield  {journal} {\bibinfo
  {journal} {Rev. Sci. Instrum.}\ }\textbf {\bibinfo {volume} {79}} (\bibinfo
  {year} {2008})}\BibitemShut {NoStop}%
\bibitem [{\citenamefont {Bai}\ \emph {et~al.}(2020)\citenamefont {Bai},
  \citenamefont {Zhou}, \citenamefont {Wang}, \citenamefont {Wu}, \citenamefont
  {McGilly}, \citenamefont {Halbertal}, \citenamefont {Lo}, \citenamefont
  {Liu}, \citenamefont {Ardelean}, \citenamefont {Rivera}, \citenamefont
  {Finney}, \citenamefont {Yang}, \citenamefont {Basov}, \citenamefont {Yao},
  \citenamefont {Xu}, \citenamefont {Hone}, \citenamefont {Pasupathy},\ and\
  \citenamefont {Zhu}}]{bai2020excitons}%
  \BibitemOpen
  \bibfield  {author} {\bibinfo {author} {\bibfnamefont {Y.}~\bibnamefont
  {Bai}}, \bibinfo {author} {\bibfnamefont {L.}~\bibnamefont {Zhou}}, \bibinfo
  {author} {\bibfnamefont {J.}~\bibnamefont {Wang}}, \bibinfo {author}
  {\bibfnamefont {W.}~\bibnamefont {Wu}}, \bibinfo {author} {\bibfnamefont
  {L.~J.}\ \bibnamefont {McGilly}}, \bibinfo {author} {\bibfnamefont
  {D.}~\bibnamefont {Halbertal}}, \bibinfo {author} {\bibfnamefont {C.~F.~B.}\
  \bibnamefont {Lo}}, \bibinfo {author} {\bibfnamefont {F.}~\bibnamefont
  {Liu}}, \bibinfo {author} {\bibfnamefont {J.}~\bibnamefont {Ardelean}},
  \bibinfo {author} {\bibfnamefont {P.}~\bibnamefont {Rivera}}, \bibinfo
  {author} {\bibfnamefont {N.~R.}\ \bibnamefont {Finney}}, \bibinfo {author}
  {\bibfnamefont {X.-C.}\ \bibnamefont {Yang}}, \bibinfo {author}
  {\bibfnamefont {D.~N.}\ \bibnamefont {Basov}}, \bibinfo {author}
  {\bibfnamefont {W.}~\bibnamefont {Yao}}, \bibinfo {author} {\bibfnamefont
  {X.}~\bibnamefont {Xu}}, \bibinfo {author} {\bibfnamefont {J.}~\bibnamefont
  {Hone}}, \bibinfo {author} {\bibfnamefont {A.~N.}\ \bibnamefont
  {Pasupathy}},\ and\ \bibinfo {author} {\bibfnamefont {X.-Y.}\ \bibnamefont
  {Zhu}},\ }\bibfield  {title} {\bibinfo {title} {Excitons in strain-induced
  one-dimensional moir\'e potentials at transition metal dichalcogenide
  heterojunctions},\ }\href@noop {} {\bibfield  {journal} {\bibinfo  {journal}
  {Nat. Mater.}\ }\textbf {\bibinfo {volume} {19}},\ \bibinfo {pages} {1068}
  (\bibinfo {year} {2020})}\BibitemShut {NoStop}%
\bibitem [{\citenamefont {Wang}\ \emph {et~al.}(2021)\citenamefont {Wang},
  \citenamefont {Shi}, \citenamefont {Shih}, \citenamefont {Zhou},
  \citenamefont {Wu}, \citenamefont {Bai}, \citenamefont {Rhodes},
  \citenamefont {Barmak}, \citenamefont {Hone}, \citenamefont {Dean},\ and\
  \citenamefont {Zhu}}]{wang2021diffusivity}%
  \BibitemOpen
  \bibfield  {author} {\bibinfo {author} {\bibfnamefont {J.}~\bibnamefont
  {Wang}}, \bibinfo {author} {\bibfnamefont {Q.}~\bibnamefont {Shi}}, \bibinfo
  {author} {\bibfnamefont {E.-M.}\ \bibnamefont {Shih}}, \bibinfo {author}
  {\bibfnamefont {L.}~\bibnamefont {Zhou}}, \bibinfo {author} {\bibfnamefont
  {W.}~\bibnamefont {Wu}}, \bibinfo {author} {\bibfnamefont {Y.}~\bibnamefont
  {Bai}}, \bibinfo {author} {\bibfnamefont {D.}~\bibnamefont {Rhodes}},
  \bibinfo {author} {\bibfnamefont {K.}~\bibnamefont {Barmak}}, \bibinfo
  {author} {\bibfnamefont {J.}~\bibnamefont {Hone}}, \bibinfo {author}
  {\bibfnamefont {C.~R.}\ \bibnamefont {Dean}},\ and\ \bibinfo {author}
  {\bibfnamefont {X.-Y.}\ \bibnamefont {Zhu}},\ }\bibfield  {title} {\bibinfo
  {title} {Diffusivity reveals three distinct phases of interlayer excitons in
  {${\mathrm{MoSe}}_2/{\mathrm{ {WS}e}}_2$} heterobilayers},\ }\href@noop {}
  {\bibfield  {journal} {\bibinfo  {journal} {Phys. Rev. Lett.}\ }\textbf
  {\bibinfo {volume} {126}} (\bibinfo {year} {2021})}\BibitemShut {NoStop}%
\bibitem [{\citenamefont {Adrian}\ \emph {et~al.}(2016)\citenamefont {Adrian},
  \citenamefont {Senftleben}, \citenamefont {Morgenstern},\ and\ \citenamefont
  {Baumert}}]{adrian2016complete}%
  \BibitemOpen
  \bibfield  {author} {\bibinfo {author} {\bibfnamefont {M.}~\bibnamefont
  {Adrian}}, \bibinfo {author} {\bibfnamefont {A.}~\bibnamefont {Senftleben}},
  \bibinfo {author} {\bibfnamefont {S.}~\bibnamefont {Morgenstern}},\ and\
  \bibinfo {author} {\bibfnamefont {T.}~\bibnamefont {Baumert}},\ }\bibfield
  {title} {\bibinfo {title} {Complete analysis of a transmission electron
  diffraction pattern of a {Mo{S$_2$}}--graphite heterostructure},\ }\href@noop
  {} {\bibfield  {journal} {\bibinfo  {journal} {Ultramicroscopy}\ }\textbf
  {\bibinfo {volume} {166}},\ \bibinfo {pages} {9} (\bibinfo {year}
  {2016})}\BibitemShut {NoStop}%
\bibitem [{\citenamefont {Domke}\ \emph {et~al.}(2012)\citenamefont {Domke},
  \citenamefont {Rapp}, \citenamefont {Schmidt},\ and\ \citenamefont
  {Huber}}]{domke2012ultrafast}%
  \BibitemOpen
  \bibfield  {author} {\bibinfo {author} {\bibfnamefont {M.}~\bibnamefont
  {Domke}}, \bibinfo {author} {\bibfnamefont {S.}~\bibnamefont {Rapp}},
  \bibinfo {author} {\bibfnamefont {M.}~\bibnamefont {Schmidt}},\ and\ \bibinfo
  {author} {\bibfnamefont {H.~P.}\ \bibnamefont {Huber}},\ }\bibfield  {title}
  {\bibinfo {title} {Ultrafast pump-probe microscopy with high temporal dynamic
  range},\ }\href@noop {} {\bibfield  {journal} {\bibinfo  {journal} {Opt.
  Express}\ }\textbf {\bibinfo {volume} {20}},\ \bibinfo {pages} {10330}
  (\bibinfo {year} {2012})}\BibitemShut {NoStop}%
\bibitem [{\citenamefont {Disa}\ \emph {et~al.}(2020)\citenamefont {Disa},
  \citenamefont {Fechner}, \citenamefont {Nova}, \citenamefont {Liu},
  \citenamefont {F{\"o}rst}, \citenamefont {Prabhakaran}, \citenamefont
  {Radaelli},\ and\ \citenamefont {Cavalleri}}]{disa2020polarizing}%
  \BibitemOpen
  \bibfield  {author} {\bibinfo {author} {\bibfnamefont {A.~S.}\ \bibnamefont
  {Disa}}, \bibinfo {author} {\bibfnamefont {M.}~\bibnamefont {Fechner}},
  \bibinfo {author} {\bibfnamefont {T.~F.}\ \bibnamefont {Nova}}, \bibinfo
  {author} {\bibfnamefont {B.}~\bibnamefont {Liu}}, \bibinfo {author}
  {\bibfnamefont {M.}~\bibnamefont {F{\"o}rst}}, \bibinfo {author}
  {\bibfnamefont {D.}~\bibnamefont {Prabhakaran}}, \bibinfo {author}
  {\bibfnamefont {P.~G.}\ \bibnamefont {Radaelli}},\ and\ \bibinfo {author}
  {\bibfnamefont {A.}~\bibnamefont {Cavalleri}},\ }\bibfield  {title} {\bibinfo
  {title} {Polarizing an antiferromagnet by optical engineering of the crystal
  field},\ }\href@noop {} {\bibfield  {journal} {\bibinfo  {journal} {Nat.
  Phys.}\ }\textbf {\bibinfo {volume} {16}},\ \bibinfo {pages} {937} (\bibinfo
  {year} {2020})}\BibitemShut {NoStop}%
\bibitem [{\citenamefont {Philipp}\ \emph {et~al.}(2022)\citenamefont
  {Philipp}, \citenamefont {Tate}, \citenamefont {Shanks}, \citenamefont
  {Mele}, \citenamefont {Peemen}, \citenamefont {Dona}, \citenamefont
  {Hartong}, \citenamefont {van Veen}, \citenamefont {Shao}, \citenamefont
  {Chen} \emph {et~al.}}]{philipp2022very}%
  \BibitemOpen
  \bibfield  {author} {\bibinfo {author} {\bibfnamefont {H.~T.}\ \bibnamefont
  {Philipp}}, \bibinfo {author} {\bibfnamefont {M.~W.}\ \bibnamefont {Tate}},
  \bibinfo {author} {\bibfnamefont {K.~S.}\ \bibnamefont {Shanks}}, \bibinfo
  {author} {\bibfnamefont {L.}~\bibnamefont {Mele}}, \bibinfo {author}
  {\bibfnamefont {M.}~\bibnamefont {Peemen}}, \bibinfo {author} {\bibfnamefont
  {P.}~\bibnamefont {Dona}}, \bibinfo {author} {\bibfnamefont {R.}~\bibnamefont
  {Hartong}}, \bibinfo {author} {\bibfnamefont {G.}~\bibnamefont {van Veen}},
  \bibinfo {author} {\bibfnamefont {Y.-T.}\ \bibnamefont {Shao}}, \bibinfo
  {author} {\bibfnamefont {Z.}~\bibnamefont {Chen}}, \emph {et~al.},\
  }\bibfield  {title} {\bibinfo {title} {Very-high dynamic range, 10,000
  frames/second pixel array detector for electron microscopy},\ }\href
  {https://doi.org/10.1017/S1431927622000174} {\bibfield  {journal} {\bibinfo
  {journal} {Microsc. Microanal.}\ }\textbf {\bibinfo {volume} {28}},\ \bibinfo
  {pages} {425} (\bibinfo {year} {2022})}\BibitemShut {NoStop}%
\bibitem [{\citenamefont {Kim}\ \emph {et~al.}(2022)\citenamefont {Kim},
  \citenamefont {Ko}, \citenamefont {Jo}, \citenamefont {Kim}, \citenamefont
  {Yoo}, \citenamefont {Son},\ and\ \citenamefont {Cheong}}]{kim2022anomalous}%
  \BibitemOpen
  \bibfield  {author} {\bibinfo {author} {\bibfnamefont {J.}~\bibnamefont
  {Kim}}, \bibinfo {author} {\bibfnamefont {E.}~\bibnamefont {Ko}}, \bibinfo
  {author} {\bibfnamefont {J.}~\bibnamefont {Jo}}, \bibinfo {author}
  {\bibfnamefont {M.}~\bibnamefont {Kim}}, \bibinfo {author} {\bibfnamefont
  {H.}~\bibnamefont {Yoo}}, \bibinfo {author} {\bibfnamefont {Y.-W.}\
  \bibnamefont {Son}},\ and\ \bibinfo {author} {\bibfnamefont {H.}~\bibnamefont
  {Cheong}},\ }\bibfield  {title} {\bibinfo {title} {Anomalous optical
  excitations from arrays of whirlpooled lattice distortions in moir\'e
  superlattices},\ }\href@noop {} {\bibfield  {journal} {\bibinfo  {journal}
  {Nat. Mater.}\ } (\bibinfo {year} {2022})}\BibitemShut {NoStop}%
\bibitem [{\citenamefont {Kealhofer}\ \emph {et~al.}(2015)\citenamefont
  {Kealhofer}, \citenamefont {Lahme}, \citenamefont {Urban},\ and\
  \citenamefont {Baum}}]{kealhofer2015signal}%
  \BibitemOpen
  \bibfield  {author} {\bibinfo {author} {\bibfnamefont {C.}~\bibnamefont
  {Kealhofer}}, \bibinfo {author} {\bibfnamefont {S.}~\bibnamefont {Lahme}},
  \bibinfo {author} {\bibfnamefont {T.}~\bibnamefont {Urban}},\ and\ \bibinfo
  {author} {\bibfnamefont {P.}~\bibnamefont {Baum}},\ }\bibfield  {title}
  {\bibinfo {title} {Signal-to-noise in femtosecond electron diffraction},\
  }\href@noop {} {\bibfield  {journal} {\bibinfo  {journal} {Ultramicroscopy}\
  }\textbf {\bibinfo {volume} {159}},\ \bibinfo {pages} {19} (\bibinfo {year}
  {2015})}\BibitemShut {NoStop}%
\bibitem [{\citenamefont {Chase}\ \emph {et~al.}(2016)\citenamefont {Chase},
  \citenamefont {Trigo}, \citenamefont {Reid}, \citenamefont {Li},
  \citenamefont {Vecchione}, \citenamefont {Shen}, \citenamefont {Weathersby},
  \citenamefont {Coffee}, \citenamefont {Hartmann}, \citenamefont {Reis} \emph
  {et~al.}}]{chase2016ultrafast}%
  \BibitemOpen
  \bibfield  {author} {\bibinfo {author} {\bibfnamefont {T.}~\bibnamefont
  {Chase}}, \bibinfo {author} {\bibfnamefont {M.}~\bibnamefont {Trigo}},
  \bibinfo {author} {\bibfnamefont {A.}~\bibnamefont {Reid}}, \bibinfo {author}
  {\bibfnamefont {R.}~\bibnamefont {Li}}, \bibinfo {author} {\bibfnamefont
  {T.}~\bibnamefont {Vecchione}}, \bibinfo {author} {\bibfnamefont
  {X.}~\bibnamefont {Shen}}, \bibinfo {author} {\bibfnamefont {S.}~\bibnamefont
  {Weathersby}}, \bibinfo {author} {\bibfnamefont {R.}~\bibnamefont {Coffee}},
  \bibinfo {author} {\bibfnamefont {N.}~\bibnamefont {Hartmann}}, \bibinfo
  {author} {\bibfnamefont {D.}~\bibnamefont {Reis}}, \emph {et~al.},\
  }\bibfield  {title} {\bibinfo {title} {Ultrafast electron diffraction from
  non-equilibrium phonons in femtosecond laser heated au films},\ }\href@noop
  {} {\bibfield  {journal} {\bibinfo  {journal} {Appl. Phys. Lett.}\ }\textbf
  {\bibinfo {volume} {108}} (\bibinfo {year} {2016})}\BibitemShut {NoStop}%
\bibitem [{\citenamefont {Yang}\ \emph {et~al.}(2020)\citenamefont {Yang},
  \citenamefont {Zhu}, \citenamefont {F.~Nunes}, \citenamefont {Yu},
  \citenamefont {Parrish}, \citenamefont {Wolf}, \citenamefont {Centurion},
  \citenamefont {G{\"u}hr}, \citenamefont {Li}, \citenamefont {Liu} \emph
  {et~al.}}]{yang2020simultaneous}%
  \BibitemOpen
  \bibfield  {author} {\bibinfo {author} {\bibfnamefont {J.}~\bibnamefont
  {Yang}}, \bibinfo {author} {\bibfnamefont {X.}~\bibnamefont {Zhu}}, \bibinfo
  {author} {\bibfnamefont {J.~P.}\ \bibnamefont {F.~Nunes}}, \bibinfo {author}
  {\bibfnamefont {J.~K.}\ \bibnamefont {Yu}}, \bibinfo {author} {\bibfnamefont
  {R.~M.}\ \bibnamefont {Parrish}}, \bibinfo {author} {\bibfnamefont {T.~J.}\
  \bibnamefont {Wolf}}, \bibinfo {author} {\bibfnamefont {M.}~\bibnamefont
  {Centurion}}, \bibinfo {author} {\bibfnamefont {M.}~\bibnamefont {G{\"u}hr}},
  \bibinfo {author} {\bibfnamefont {R.}~\bibnamefont {Li}}, \bibinfo {author}
  {\bibfnamefont {Y.}~\bibnamefont {Liu}}, \emph {et~al.},\ }\bibfield  {title}
  {\bibinfo {title} {Simultaneous observation of nuclear and electronic
  dynamics by ultrafast electron diffraction},\ }\href@noop {} {\bibfield
  {journal} {\bibinfo  {journal} {Science}\ }\textbf {\bibinfo {volume}
  {368}},\ \bibinfo {pages} {885} (\bibinfo {year} {2020})}\BibitemShut
  {NoStop}%
\bibitem [{\citenamefont {Weiss}\ \emph {et~al.}(2017)\citenamefont {Weiss},
  \citenamefont {Shanks}, \citenamefont {Philipp}, \citenamefont {Becker},
  \citenamefont {Chamberlain}, \citenamefont {Purohit}, \citenamefont {Tate},\
  and\ \citenamefont {Gruner}}]{weiss2017high}%
  \BibitemOpen
  \bibfield  {author} {\bibinfo {author} {\bibfnamefont {J.~T.}\ \bibnamefont
  {Weiss}}, \bibinfo {author} {\bibfnamefont {K.~S.}\ \bibnamefont {Shanks}},
  \bibinfo {author} {\bibfnamefont {H.~T.}\ \bibnamefont {Philipp}}, \bibinfo
  {author} {\bibfnamefont {J.}~\bibnamefont {Becker}}, \bibinfo {author}
  {\bibfnamefont {D.}~\bibnamefont {Chamberlain}}, \bibinfo {author}
  {\bibfnamefont {P.}~\bibnamefont {Purohit}}, \bibinfo {author} {\bibfnamefont
  {M.~W.}\ \bibnamefont {Tate}},\ and\ \bibinfo {author} {\bibfnamefont
  {S.~M.}\ \bibnamefont {Gruner}},\ }\bibfield  {title} {\bibinfo {title} {High
  dynamic range x-ray detector pixel architectures utilizing charge removal},\
  }\href@noop {} {\bibfield  {journal} {\bibinfo  {journal} {IEEE Trans. Nucl.
  Sci.}\ }\textbf {\bibinfo {volume} {64}},\ \bibinfo {pages} {1101} (\bibinfo
  {year} {2017})}\BibitemShut {NoStop}%
\bibitem [{\citenamefont {Muller}\ \emph {et~al.}(2006)\citenamefont {Muller},
  \citenamefont {Kirkland}, \citenamefont {Thomas}, \citenamefont {Grazul},
  \citenamefont {Fitting},\ and\ \citenamefont {Weyland}}]{muller2006room}%
  \BibitemOpen
  \bibfield  {author} {\bibinfo {author} {\bibfnamefont {D.~A.}\ \bibnamefont
  {Muller}}, \bibinfo {author} {\bibfnamefont {E.~J.}\ \bibnamefont
  {Kirkland}}, \bibinfo {author} {\bibfnamefont {M.~G.}\ \bibnamefont
  {Thomas}}, \bibinfo {author} {\bibfnamefont {J.~L.}\ \bibnamefont {Grazul}},
  \bibinfo {author} {\bibfnamefont {L.}~\bibnamefont {Fitting}},\ and\ \bibinfo
  {author} {\bibfnamefont {M.}~\bibnamefont {Weyland}},\ }\bibfield  {title}
  {\bibinfo {title} {Room design for high-performance electron microscopy},\
  }\href@noop {} {\bibfield  {journal} {\bibinfo  {journal} {Ultramicroscopy}\
  }\textbf {\bibinfo {volume} {106}},\ \bibinfo {pages} {1033} (\bibinfo {year}
  {2006})}\BibitemShut {NoStop}%
\bibitem [{\citenamefont {Chatelain}\ \emph {et~al.}(2014)\citenamefont
  {Chatelain}, \citenamefont {Morrison}, \citenamefont {Klarenaar},\ and\
  \citenamefont {Siwick}}]{chatelain2014coherent}%
  \BibitemOpen
  \bibfield  {author} {\bibinfo {author} {\bibfnamefont {R.~P.}\ \bibnamefont
  {Chatelain}}, \bibinfo {author} {\bibfnamefont {V.~R.}\ \bibnamefont
  {Morrison}}, \bibinfo {author} {\bibfnamefont {B.~L.~M.}\ \bibnamefont
  {Klarenaar}},\ and\ \bibinfo {author} {\bibfnamefont {B.~J.}\ \bibnamefont
  {Siwick}},\ }\bibfield  {title} {\bibinfo {title} {Coherent and incoherent
  electron-phonon coupling in graphite observed with radio-frequency compressed
  ultrafast electron diffraction},\ }\href@noop {} {\bibfield  {journal}
  {\bibinfo  {journal} {Phys. Rev. Lett.}\ }\textbf {\bibinfo {volume} {113}}
  (\bibinfo {year} {2014})}\BibitemShut {NoStop}%
\bibitem [{\citenamefont {Zalden}\ \emph {et~al.}(2019)\citenamefont {Zalden},
  \citenamefont {Quirin}, \citenamefont {Schumacher}, \citenamefont {Siegel},
  \citenamefont {Wei}, \citenamefont {Koc}, \citenamefont {Nicoul},
  \citenamefont {Trigo}, \citenamefont {Andreasson}, \citenamefont {Enquist},
  \citenamefont {Shu}, \citenamefont {Pardini}, \citenamefont {Chollet},
  \citenamefont {Zhu}, \citenamefont {Lemke}, \citenamefont {Ronneberger},
  \citenamefont {Larsson}, \citenamefont {Lindenberg}, \citenamefont {Fischer},
  \citenamefont {Hau-Riege}, \citenamefont {Reis}, \citenamefont {Mazzarello},
  \citenamefont {Wuttig},\ and\ \citenamefont
  {Sokolowski-Tinten}}]{zalden2019femtosecond}%
  \BibitemOpen
  \bibfield  {author} {\bibinfo {author} {\bibfnamefont {P.}~\bibnamefont
  {Zalden}}, \bibinfo {author} {\bibfnamefont {F.}~\bibnamefont {Quirin}},
  \bibinfo {author} {\bibfnamefont {M.}~\bibnamefont {Schumacher}}, \bibinfo
  {author} {\bibfnamefont {J.}~\bibnamefont {Siegel}}, \bibinfo {author}
  {\bibfnamefont {S.}~\bibnamefont {Wei}}, \bibinfo {author} {\bibfnamefont
  {A.}~\bibnamefont {Koc}}, \bibinfo {author} {\bibfnamefont {M.}~\bibnamefont
  {Nicoul}}, \bibinfo {author} {\bibfnamefont {M.}~\bibnamefont {Trigo}},
  \bibinfo {author} {\bibfnamefont {P.}~\bibnamefont {Andreasson}}, \bibinfo
  {author} {\bibfnamefont {H.}~\bibnamefont {Enquist}}, \bibinfo {author}
  {\bibfnamefont {M.~J.}\ \bibnamefont {Shu}}, \bibinfo {author} {\bibfnamefont
  {T.}~\bibnamefont {Pardini}}, \bibinfo {author} {\bibfnamefont
  {M.}~\bibnamefont {Chollet}}, \bibinfo {author} {\bibfnamefont
  {D.}~\bibnamefont {Zhu}}, \bibinfo {author} {\bibfnamefont {H.}~\bibnamefont
  {Lemke}}, \bibinfo {author} {\bibfnamefont {I.}~\bibnamefont {Ronneberger}},
  \bibinfo {author} {\bibfnamefont {J.}~\bibnamefont {Larsson}}, \bibinfo
  {author} {\bibfnamefont {A.~M.}\ \bibnamefont {Lindenberg}}, \bibinfo
  {author} {\bibfnamefont {H.~E.}\ \bibnamefont {Fischer}}, \bibinfo {author}
  {\bibfnamefont {S.}~\bibnamefont {Hau-Riege}}, \bibinfo {author}
  {\bibfnamefont {D.~A.}\ \bibnamefont {Reis}}, \bibinfo {author}
  {\bibfnamefont {R.}~\bibnamefont {Mazzarello}}, \bibinfo {author}
  {\bibfnamefont {M.}~\bibnamefont {Wuttig}},\ and\ \bibinfo {author}
  {\bibfnamefont {K.}~\bibnamefont {Sokolowski-Tinten}},\ }\bibfield  {title}
  {\bibinfo {title} {Femtosecond x-ray diffraction reveals a liquid -- liquid
  phase transition in phase-change materials},\ }\href@noop {} {\bibfield
  {journal} {\bibinfo  {journal} {Science}\ }\textbf {\bibinfo {volume}
  {364}},\ \bibinfo {pages} {1062} (\bibinfo {year} {2019})}\BibitemShut
  {NoStop}%
\bibitem [{\citenamefont {Otto}\ \emph {et~al.}(2017)\citenamefont {Otto},
  \citenamefont {Ren{\'e}~de Cotret}, \citenamefont {Stern},\ and\
  \citenamefont {Siwick}}]{otto2017solving}%
  \BibitemOpen
  \bibfield  {author} {\bibinfo {author} {\bibfnamefont {M.~R.}\ \bibnamefont
  {Otto}}, \bibinfo {author} {\bibfnamefont {L.}~\bibnamefont {Ren{\'e}~de
  Cotret}}, \bibinfo {author} {\bibfnamefont {M.~J.}\ \bibnamefont {Stern}},\
  and\ \bibinfo {author} {\bibfnamefont {B.~J.}\ \bibnamefont {Siwick}},\
  }\bibfield  {title} {\bibinfo {title} {Solving the jitter problem in
  microwave compressed ultrafast electron diffraction instruments: {R}obust
  sub-50 fs cavity-laser phase stabilization},\ }\href@noop {} {\bibfield
  {journal} {\bibinfo  {journal} {Struct. Dyn.}\ }\textbf {\bibinfo {volume}
  {4}} (\bibinfo {year} {2017})}\BibitemShut {NoStop}%
\bibitem [{\citenamefont {Weiss}\ \emph {et~al.}(2016)\citenamefont {Weiss},
  \citenamefont {Becker}, \citenamefont {Shanks}, \citenamefont {Philipp},
  \citenamefont {Tate},\ and\ \citenamefont {Gruner}}]{weiss2016potential}%
  \BibitemOpen
  \bibfield  {author} {\bibinfo {author} {\bibfnamefont {J.~T.}\ \bibnamefont
  {Weiss}}, \bibinfo {author} {\bibfnamefont {J.}~\bibnamefont {Becker}},
  \bibinfo {author} {\bibfnamefont {K.~S.}\ \bibnamefont {Shanks}}, \bibinfo
  {author} {\bibfnamefont {H.~T.}\ \bibnamefont {Philipp}}, \bibinfo {author}
  {\bibfnamefont {M.~W.}\ \bibnamefont {Tate}},\ and\ \bibinfo {author}
  {\bibfnamefont {S.~M.}\ \bibnamefont {Gruner}},\ }\bibfield  {title}
  {\bibinfo {title} {Potential beneficial effects of electron-hole plasmas
  created in silicon sensors by { {XFEL}}-like high intensity pulses for
  detector development},\ }\href@noop {} {\bibfield  {journal} {\bibinfo
  {journal} {AIP Conf. Proc.}\ }\textbf {\bibinfo {volume} {1741}} (\bibinfo
  {year} {2016})}\BibitemShut {NoStop}%
\bibitem [{\citenamefont {Schneider}\ \emph {et~al.}(2010)\citenamefont
  {Schneider}, \citenamefont {Calado}, \citenamefont {Zandbergen},
  \citenamefont {Vandersypen},\ and\ \citenamefont
  {Dekker}}]{schneider2010wedging}%
  \BibitemOpen
  \bibfield  {author} {\bibinfo {author} {\bibfnamefont {G.~F.}\ \bibnamefont
  {Schneider}}, \bibinfo {author} {\bibfnamefont {V.~E.}\ \bibnamefont
  {Calado}}, \bibinfo {author} {\bibfnamefont {H.}~\bibnamefont {Zandbergen}},
  \bibinfo {author} {\bibfnamefont {L.~M.~K.}\ \bibnamefont {Vandersypen}},\
  and\ \bibinfo {author} {\bibfnamefont {C.}~\bibnamefont {Dekker}},\
  }\bibfield  {title} {\bibinfo {title} {Wedging transfer of nanostructures},\
  }\href@noop {} {\bibfield  {journal} {\bibinfo  {journal} {Nano Lett.}\
  }\textbf {\bibinfo {volume} {10}},\ \bibinfo {pages} {1912} (\bibinfo {year}
  {2010})}\BibitemShut {NoStop}%
\bibitem [{\citenamefont {Zhang}\ and\ \citenamefont
  {Tadmor}(2018)}]{zhang2018structural}%
  \BibitemOpen
  \bibfield  {author} {\bibinfo {author} {\bibfnamefont {K.}~\bibnamefont
  {Zhang}}\ and\ \bibinfo {author} {\bibfnamefont {E.~B.}\ \bibnamefont
  {Tadmor}},\ }\bibfield  {title} {\bibinfo {title} {Structural and electron
  diffraction scaling of twisted graphene bilayers},\ }\href@noop {} {\bibfield
   {journal} {\bibinfo  {journal} {J. Mech. Phys. Solids}\ }\textbf {\bibinfo
  {volume} {112}},\ \bibinfo {pages} {225} (\bibinfo {year}
  {2018})}\BibitemShut {NoStop}%
\end{thebibliography}
\end{document}